\documentclass[%
 reprint,
nofootinbib,
 amsmath,amssymb,
 aps,
pre
]{revtex4-2}

\usepackage{graphicx}
\usepackage{dcolumn}
\usepackage{bm}
\usepackage[english]{babel}
\usepackage{upgreek}
\usepackage{subcaption}

\usepackage{comment}

\newcommand{\mc}[1]{\mathcal{#1}}

\newcommand{\half}[0]{\frac{1}{2}}

\newcommand{\paren}[1]{\left(#1\right)}

\newcommand{\consfrac}[3]{\left(\frac{\partial #1}{\partial #2}\right)_{#3}}

\newcommand{\zero}[0]{{(0)}}
\newcommand{\one}[0]{{(1)}}

\newcommand{\si}{\sigma}
\newcommand{\eps}{\epsilon}
\newcommand{\dtail}{\eta} 
\newcommand{\symbdTdp}{\rho} 
\newcommand{\Scloud}{\xi} 
\newcommand{\ph}{{g_i}} 
\newcommand{\php}{{g_i'}} 
\newcommand{\kL}{\Lambda} 
\newcommand{\Q}{Q} 
\newcommand{\heatcap}{{c_p}} 

\newcommand{\eq}{^{\text{(eq)}}}

\newcommand{\pcset}{\mc P}
\newcommand{\pcrit}{p^*}
\newcommand{\ch}{{c-h}}
\newcommand{\hc}{{h-c}}

\begin{document}

\preprint{APS/123-QED}

\title{Evolution of the Radiative Thermal Instability in a Confined Medium}

\author{Henry Fetsch}
 \email{hfetsch@princeton.edu}
\author{Nathaniel J. Fisch}%
\affiliation{%
 Department of Astrophysical Sciences\\
 Princeton University, Princeton, NJ 08540
}%

\date{\today}

\begin{abstract}

Thermally bistable fluid tends to self-organize into clouds of hot and cold material, which are internally uniform and separated by thin conduction fronts. The evolution of these clouds has been studied for isobaric systems, but when pressure is instead treated as a dynamical quantity and allowed to evolve self-consistently, fundamentally different dynamics appear. Such a treatment is necessary in some laboratory plasmas, whose volume is constrained but whose pressure can vary. Solutions are derived for the evolution of clouds, accounting for pressure variation and interactions between conduction fronts. Additional stable configurations and secondary instabilities are derived, which may be relevant to fusion plasmas and to the study of photoionized plasma in the laboratory.

\end{abstract}

\maketitle


\section{Introduction}
\label{sec_intro}

A thermally unstable system is one in which a positive (negative) temperature perturbation leads to further heating (cooling). The radiative thermal instability, produced by a balance between absorbed and emitted radiation, appears in a diverse array of plasma systems. In astrophysical and solar plasmas, thermal instabilities contribute to structure formation in the interstellar and intercluster medium \cite{Field_1965, Inoue_Inutsuka_Koyama_2006, ZuHone_Roediger_2016} and in the dynamics of solar prominences \cite{Field_1965, Soler_Ballester_Goossens_2011, Somov_Syrovatskii_1982, Nakagawa_1970}. Thermal instabilities can degrade--or in some cases improve--the confinement of laboratory fusion plasma, appearing notably as `MARFES' in tokamak edge regions \cite{ASDEX_2023, Tokar_2002, Stroth_Bernert_Brida_Cavedon_Dux_Huett_Lunt_Pan_Wischmeier_Team_2022, Bernert_et_2023} and as strongly radiating filaments in z-pinches \cite{Haines_1989}, and play a role in a variety of other laboratory experiments \cite{Moore_Gumbrell_Lazarus_Hohenberger_Robinson_Smith_Plant_Symes_Dunne_2008, Suzuki-Vidal_et_2015}.

If a thermally unstable fluid is initially in equilibrium, then a spatially nonuniform perturbation will evolve into a patchwork of hot and cold regions \cite{Field_1965, Meerson_1989, Inutsuka_Koyama_Inoue_2005}. A thermally bistable fluid admits two stable temperatures, one hotter and one colder than the unstable initial state. As the linear instability saturates, 
most fluid elements reach stable equilibria locally, excepting narrow transition layers between hot and cold regions.

On larger scales, such a system can be viewed as a two-phase medium with a hot phase at temperature $T_h$ and a cold phase at temperature $T_c$. In the boundary layers between phases, termed conduction fronts, thermal conduction becomes important. For isolated planar fronts, the propagation speed, either into the hot phase (condensation) or into the cold phase (evaporation), depends only on the system pressure. There exists a critical pressure $\pcrit$ at which a conduction front is stationary, assuming that it separates semi-infinite regions at temperatures $T_h$ and $T_c$  \cite{ZP69, Penston_Brown_1970}. For systems consisting of multiple fronts, the dynamics are more complicated. The nonlinear nature of the governing equations leads to interactions between nearby fronts; at critical pressure, this generally causes the collapse of small-scale features and the generation of large-scale structure \cite{Elphick_Regev_Spiegel_1991, Elphick_Regev_Shaviv_1992}.

Although further complications are introduced by the curvature of fronts, we isolate the effects of pressure variations by considering one-dimensional systems in this work. Considerable complexity and interesting phenomena are retained even under this simplification \cite{ZP69, Elphick_Regev_Spiegel_1991, Aranson_Meerson_Sasorov_1993}. In magnetized plasma, dynamics in the parallel direction often dominate, naturally prompting a one-dimensional approximation \cite{Field_1965}.

Analytical treatments of conduction fronts in thermally bistable fluid have typically treated pressure as a fixed quantity, meaning that the whole system must expand or contract when the average temperature changes. In systems at fixed volume, as is the case in some experiments, a different (isochoric) constraint should be applied. Pressure must then be treated as a dynamical quantity. Few past studies have taken this step; a notable exception is the work of Aranson, Meerson, and Sasorov \cite{Aranson_Meerson_Sasorov_1993} (hereafter AMS), who considered a thermally bistable system in a fixed volume. They derived a relation (corresponding to Eq.~\eqref{eq_chi_defn} in this work) showing how the motion of conduction fronts causes system pressure to change. 
Our work generalizes these results in several ways, providing an analysis that applies at later times and to more general classes of cooling functions. The question considered here is how such systems evolve at late times, when hot and cold regions have fully developed and form, in essence, a cloudy mixture.

The classical theory of thermal instability deals with the short-timescale linear response to temperature perturbations; if the initial state is unstable, perturbations grow. Growth of the primary instability saturates when each region arrives at the stable cold ($T_c$) or hot ($T_h$) phase. At this point, the system is generally not in a true steady state, but instead exhibits a slow evolution mode driven by conduction. If pressure is constant, then few steady states are possible and the system becomes homogeneous at late times. If pressure is dynamical, then it can be viewed as an additional degree of freedom; this work explores the consequences of this degree of freedom. For parameters of interest, pressure remains spatially uniform but evolves over time. For some cooling curves, pressure variation can drive a secondary instability, rapidly reshaping the hot and cold regions. Since more of configuration space is available, the space of accessible steady states is fundamentally altered. On long timescales, interfront interactions become important and lead to steady states different from those found by AMS. In contrast to the constant-pressure case, this allows a stable, non-homogeneous steady state, which may be observable in astrophysical systems and in experiments.

As an intermediate step, we derive formulas (Eq.\eqref{eq_j1_p}, Eq.~\eqref{eq_j1_h}, and Eq.~\eqref{eq_j1_nu}) for the evolution of conduction fronts due to deviations from critical pressure, interactions between fronts, and time-varying pressure. The latter two are more general than those previously published; they involve only explicit functionals of the cooling curve and thermal conductivity, and do not require computing the profiles of the fronts. Our results therefore apply to any cooling curve, provided that a few reasonable conditions are met, rather than a specific curve with a simple form. This generality reveals nuances of front dynamics that were obfuscated by the symmetry of the cooling curves used in previous treatments.

This paper is organized as follows. In \S\ref{sec_definitions_assumptions}, we outline the system under consideration and our assumptions, and in \S\ref{sec_thermal_instability} we review the basic physics of the relevant thermal instability. Establishing common ground with previous works, we derive in \S\ref{sec_isobaric} the evolution equations for conduction fronts in a system at fixed pressure. In \S\ref{sec_isochoric}, we examine the dynamics of fronts in systems whose volume, rather than pressure, is constrained. In \S\ref{sec_evolution}, we expand this treatment to ensembles of fronts constituting a `cloudy' medium with complex spatial structure. These evolution equations become analytically intractable, so in \S\ref{sec_numerics} we present numerical solutions. Finally, in \S\ref{sec_discussion} we discuss the implications of these results for laboratory and astrophysical systems. Our demonstration that isochoric conditions promote the formation of long-lived structures in photoionized may be relevant to opacity measurements relying on assumptions of spatial uniformity.

\section{Definitions and Assumptions}
\label{sec_definitions_assumptions}

\subsection{Heating and cooling}
\label{sec_cooling}

We consider a fluid absorbing energy from some external source, and also radiating energy away, at a rate dependent on its number density $n$ and temperature $T$. Temperature has units of energy in this work. The fluid is assumed to be optically thin to both the external source and its own emission such that radiation transport can be neglected. The power flow due to absorption and radiation is described by the \textit{cooling function} $\mc L$
such that the heating and cooling of a fluid element is described by
\begin{equation}
\label{eq_dsdt_3d}
    nT\frac{d\mc S}{dt} = -\mc L + \nabla \cdot (\kappa \nabla T) ,
\end{equation}
where $\mc S$ is the entropy per particle, $d/dt$ is the convective derivative, and $\kappa$ is the thermal conductivity.

Assuming an ideal gas equation of state, the pressure $p$ is given by $p = nT$. The entropy per particle, up to an irrelevant additive constant, is ${\mc S = \ln(T^\heatcap p^{-1})}$ where $\heatcap \doteq \gamma/(\gamma-1)$ and $\gamma$ is the adiabatic index. We assume that fluid properties only vary in the $x$ direction. Combining these assumptions, Eq.~\eqref{eq_dsdt_3d} becomes
\begin{equation}
\label{eq_ddt_gen}
    \heatcap n \frac{dT}{dt} - \frac{dp}{dt} = -\mc L(p, T) + \frac{\partial}{\partial x}(\kappa \frac{\partial T}{\partial x}) ,
\end{equation}
where we have changed variables to work with pressure instead of density, and accordingly we now evaluate $\mc L$ as a function of $p$ and $T$.

\subsection{Lagrangian coordinates}
\label{sec_lagrangian_coords}

We assume that there is no background flow, so the fluid velocity $u$ will only be nonzero if the movement of conduction fronts drives flows. The effect of temperature on ionization state is neglected so that the system contains no particle sources or sinks\footnote{There are regimes in which this assumption is well justified \cite{Field_1965}; in systems where recombination and molecule formation become important, an ionization-related term would need to be added to the equations in this work, but in general the results will hold qualitatively.}. Conservation of particle number then requires
\begin{equation}
    \frac{\partial n}{\partial t} = - \frac{\partial}{\partial x}(n u).
\end{equation}

It will be convenient to adopt a coordinate system that moves with the fluid. Taking an approach that has been applied to this problem previously \cite{Balbus_Soker_1989,Meerson_1989, Elphick_Regev_Shaviv_1992}, we define a Lagrangian variable $\sigma$ representing the areal number density in the region between some reference position $x_0$ and the Eulerian position of interest $x$, viz.
\begin{equation}
    \sigma(x, t) = \int_{x_0}^x n(x', t) dx' .
\end{equation}

To convert Eq.~\eqref{eq_ddt_gen} to Lagrangian coordinates, derivatives transform as ${(d/dt) \rightarrow \partial_t}$ and ${\partial_x \rightarrow n(\sigma,t) \partial_\si}$, and so we have
\begin{equation}
\label{eq_dt_langrang_gen}
    \heatcap n\frac{\partial T}{\partial t} - \frac{\partial p}{\partial t} = -\mc L(p, t) + n\frac{\partial}{\partial\sigma}(n \kappa \frac{\partial T}{\partial\sigma}) ,
\end{equation}
where the pressure and temperature fields are now $p(\si,t)$ and $T(\si,t)$.

\subsection{Characteristic scales}
\label{sec_scales}

We denote characteristic system parameters by a subscript 0, e.g. the characteristic temperature $T_0$, density $n_0$, cooling rate $\mc L_0$, and thermal conductivity $\kappa_0$. The sound speed  is $c_s$ and the length scale $L_0$ represents the system size. Characteristic timescales include the cooling time $t_\mathrm{cl} \doteq n_0T_0/\mc L_0$ and the system sound crossing time $t_\mathrm{sys} \doteq L_0/c_s$. Conduction fronts have a characteristic width $l_0 \sim \sqrt{\kappa_0 t_\mathrm{cl}/n_0}$ and the time for sound to cross a conduction front is $t_\mathrm{fr} \doteq l_0/c_s$. We define a dimensionless parameter $\delta \doteq l_0/L_0$. Throughout this work, we assume the following ordering:
\begin{equation}
    \label{eq_timescale_ordering}
    t_\mathrm{fr} \ll t_\mathrm{sys} \ll t_\mathrm{cl} .
\end{equation}

Physically, this ordering means that conduction fronts are much smaller than the outer length scales of the system ($\delta \ll 1$) and establishment of pressure equilibrium across a front is the fastest timescale under consideration. Globally, sound waves establish pressure equilibrium throughout the system faster than external processes can significantly change the temperature of a fluid element. As a result, we may treat the pressure as spatially uniform; this is a common assumption in studies of thermal instability \cite{Elphick_Regev_Spiegel_1991, Aranson_Meerson_Sasorov_1993, Aharonson_Regev_Shaviv_1994}. We will allow pressure to vary in time as the average temperature of the system changes; as long as this change is caused by internal processes, as opposed to arbitrary external forcing, the timescale of pressure variation will be much longer than $t_\mathrm{sys}$.

\section{Thermal Instability}
\label{sec_thermal_instability}

\subsection{Linear stability}
\label{sec_linear_stability}

To describe the conduction fronts separating hot and cold phases of a thermally bistable medium requires understanding the saturated nonlinear stages of the thermal instability. We begin with a description of its linear behavior, following the seminal work of Field \cite{Field_1965}, though different notation is used.

Let us consider a general system with temperature $T$ whose evolution is governed by the function $\mc H(T)$ such that
\begin{equation}
\label{eq_dt_H_gen}
    \partial_t T = \mc H(T, Y) ,
\end{equation}
where $Y$ represents the other independent variables on which $\mc H$ depends and the subscript indicates the quantity to be held constant when evaluating the derivative.. The system is in thermal equilibrium when $\mc H = 0$. Let $T_0$ be an equilibrium temperature. The stability of the equilibrium is determined by perturbing such that $T=T_0+\widetilde T$. To linear order, the time evolution of the perturbation is determined by
\begin{equation}
    \partial_t \widetilde T = \consfrac{\mc H}{T}{Y} \widetilde T .
\end{equation}

If $\mc H^\prime(T_0) > 0$, the equilibrium is thermally unstable. To connect the general Eq.~\eqref{eq_dt_H_gen} to Eq.~\eqref{eq_dt_langrang_gen} describing our system of interest, we need an appropriate relationship between $\mc H$ and $\mc L$. Earlier work by Parker \cite{Parker53} suggested an isochoric model represented as $\mc H \propto -\mc L(n, T)$, with $\mc H^\prime \propto (\partial \mc L/\partial T)_n$ evaluated at constant $n$ (the proportionality elides positive constant factors that are unimportant for this summary).

It was subsequently shown, initially by Zanstra \cite{Zanstra_1955} and then in detail by Field \cite{Field_1965}, that attention must be paid to the adiabatic expansion and compression resulting from temperature perturbations. As a result, the heating function is best represented as $\mc H \propto -\mc L(p,T)$ such that
\begin{equation}
\label{eq_H_prime_dLdT}
    \mc H^\prime \propto - \left(\frac{\partial\mc L}{\partial T}\right)_p = -\left(\frac{\partial\mc L}{\partial T}\right)_n + \frac{n}{T}\left(\frac{\partial\mc L}{\partial n}\right)_T.
\end{equation}
For future reference, we define \begin{equation}
\label{eq_L_pT_def}
\begin{split}
    \mc L_T &\doteq \paren{\frac{\partial \mc L}{\partial T}}_p ,
    \\
    \mc L_p &\doteq \paren{\frac{\partial \mc L}{\partial p}}_T .
\end{split}
\end{equation}

A model corresponding to Eq.~\eqref{eq_H_prime_dLdT} is termed isobaric. For long-wavelength perturbations, an isochoric model becomes more accurate as the time for sound waves to establish pressure equilibrium becomes longer than the instability growth rate. For short-wavelength perturbations, thermal conduction has a stabilizing effect. These effects have been examined in detail by Field and by many others \cite{Field_1965, Meerson_1989, Balbus_Soker_1989, Inutsuka_Koyama_Inoue_2005, Inoue_Inutsuka_Koyama_2006, Illarionov_Igumenshchev_1998, Koyama_Inutsuka_2004}. Since we are not concerned in this work with the details of linear growth, we can ignore these subtleties; the ordering in Eq.~\eqref{eq_timescale_ordering} guarantees that Eq.~\eqref{eq_H_prime_dLdT} is the appropriate quantity for determining stability.

\subsection{Cooling curves}
\label{sec_cooling_curves}

Thermal instability occurs when the graph of $\mc L ~\mathrm{vs.}~T$ (at constant $p$) crosses the abscissa with negative slope; positive slope yields a stable equilibrium. For fixed pressure, let $T_w$ be the temperature of an unstable `warm' equilibrium such that ${\mc L(T_w) = 0}$ and ${\mc L_T(T_w) < 0}$. Thermodynamic consistency requires that $\mc L < 0$ as $T\rightarrow 0$ and $\mc L > 0$ as $T \rightarrow \infty$. The curve must therefore cross the abscissa in at least two other places. If there are only two other crossings, then $\mc L_T$ must be positive at both. 

\begin{figure}
    \centering
    \includegraphics[width=0.5\columnwidth]{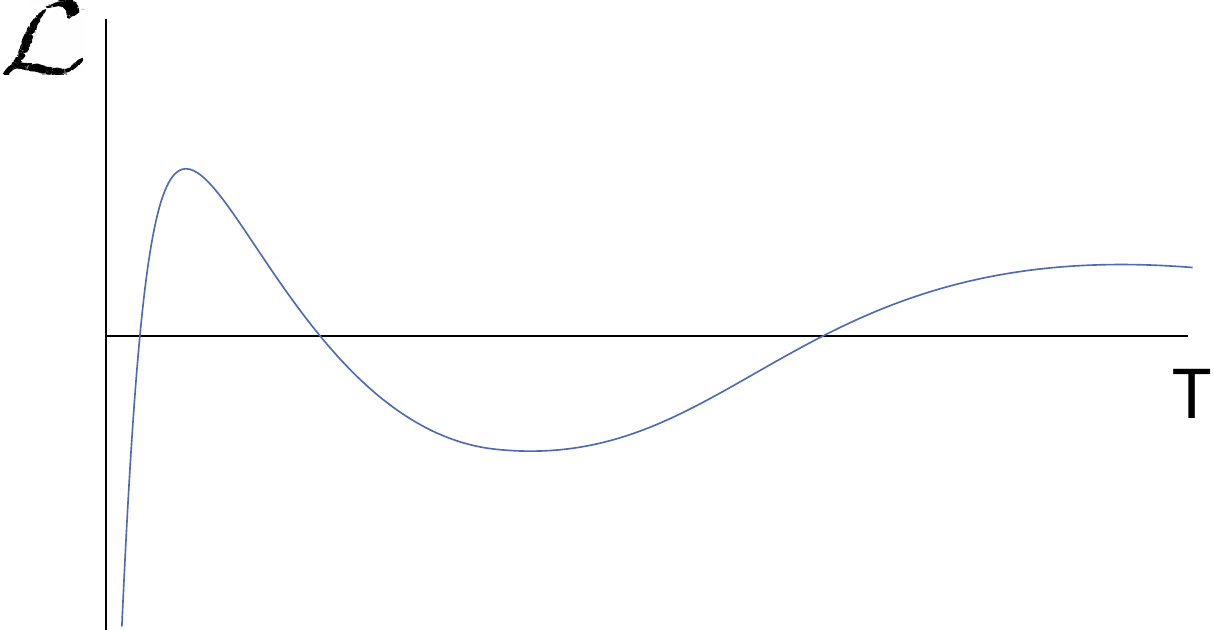}
    \caption{Cooling rate $\mc L(T)$ for a bistable cooling curve at fixed $p$.}
    \label{fig_sketch_mcL_p}
\end{figure}

We consider \textit{bistable} cooling curves, which have a single unstable equilibrium at temperature $T_w$ and two stable equilibria at temperatures $T_c$ and $T_h$. These equilibria are termed the `cold' and `hot' phases respectively. A schematic bistable $\mc L$ is shown in Fig.~\ref{fig_sketch_mcL_p}. 

The appropriate model for a cooling curve depends strongly on the medium being described and the regime of interest. For instance, in a photoionized plasma, the relative importance of bound-bound (excitation/de-excitation), bound-free (ionization/recombination) and free-free (bremsstrahlung etc.) processes changes with temperature and with density. Typically, cooling terms go as larger powers of $n$ than heating terms because emission processes usually involve more particles than absorption processes (discounting photons because they do not contribute to $n$).

\begin{figure}
    \centering
    \includegraphics[width=0.5\columnwidth]{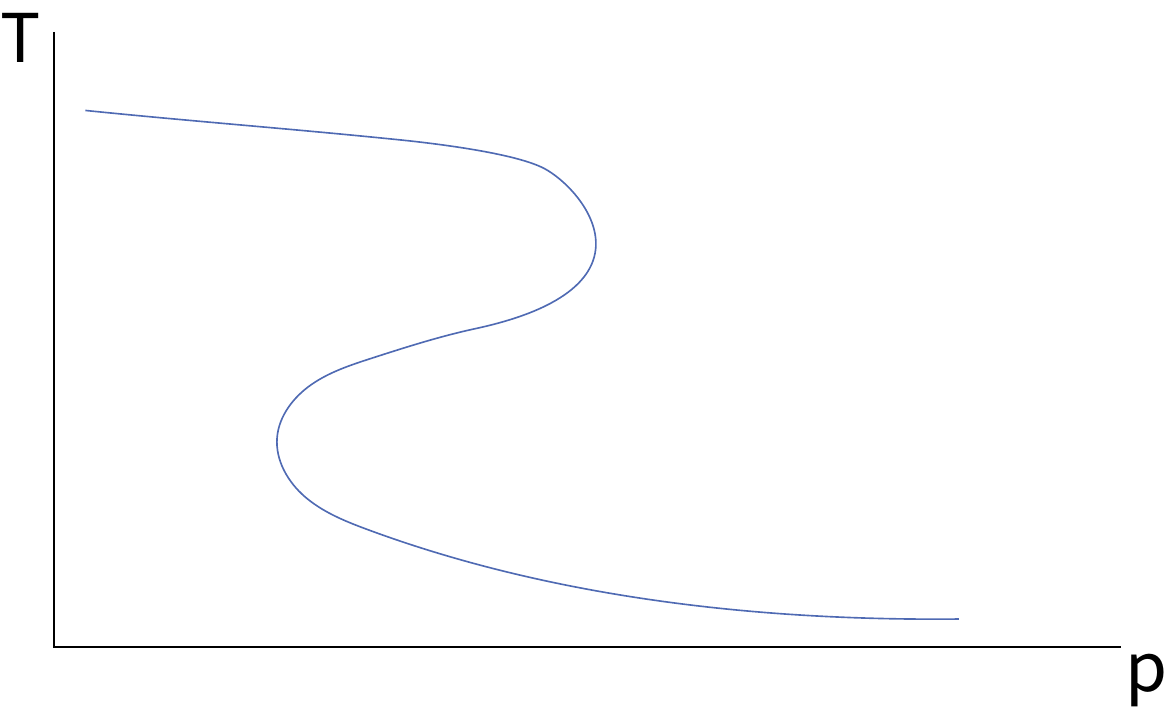}
    \caption{Zeros of the function $\mc L(p, T)$. The characteristic $S$ shape yields bistability.}
    \label{fig_sketch_T_p}
\end{figure}

For reasons described above, we choose to work with pressure rather than density, assuming as everywhere else in this work that $p=nT$. At high pressure, cooling dominates and only a single cold equilibrium is possible; at low pressure, heating dominates and we have only a single hot equilibrium; both equilibria must therefore be stable. In an intermediate regime, bistability is possible. For an example cooling curve, Fig.~\ref{fig_sketch_T_p} shows the zeros of $\mc L$ as a function of $p$ and $T$ (Fig.~\ref{fig_sketch_mcL_p} is a vertical slice from the bistable region of Fig.~\ref{fig_sketch_T_p}). The solid curve represents the zero locus of $\mc L$; every point on this curve is an equilibrium. Between the two turning points, the cooling curve is bistable.

\subsection{Conduction fronts}
\label{sec_conduction_fronts}

Analysis of cooling curves yields conditions for local thermal equilibrium, but heat conduction can destroy this equilibrium. Suppose that we set up as the initial state an arbitrary (inhomogeneous) temperature profile. Over the first few $t_\mathrm{cl}$, linear instability drives temperature everywhere toward $T_c$ or $T_h$, while heat conduction rapidly smooths temperature gradients; we make no attempt here to follow this transient process. As the transients die down, the system becomes a patchwork of hot and cold regions; deep inside each region, temperature is nearly uniform. We suppose that thermal instability is triggered everywhere such that no regions of unstable `warm' phase persist. Separating adjacent regions are narrow boundary layers, with size of order $l_0$, where conduction is important; sensibly, these layers are termed conduction fronts.

Zooming out, the details of heating and cooling in each fluid element can be abstracted away. Instead, the system can be viewed as a two-phase medium with some rule describing how the phase boundaries (conduction fronts) evolve in time. In the next sections, we will derive this rule.

In the microscopic picture the temperature profile is not perfectly flat, but in the macroscopic picture each region simply consists of either hot or cold phase. The only other relevant property of each region is its size. Since we work in Lagrangian coordinates, the `size' of a region of Eulerian width $L$ is the areal density ${\Scloud = n_gL}$ where $n_g$ is the density in the region. Motion of a front depends on the size of adjacent regions and on the system pressure.

We label fronts as $\hc$ when they connect a hot region on the left to a cold region on the right, and $\ch$ when they connect a cold region on the left to a hot region on the right. For these types of fronts, the temperature profile is monotonic in space, which will be important for variable transformations that we will perform. For fronts connecting to warm regions, the temperature profile can be oscillatory \cite{Elphick_Regev_Shaviv_1992}, but we do not consider this type of front because we are interested in long-timescale behavior. To measure the location of and distance between fronts, we define the location of a front to be the point $\sigma$ (in Lagrangian coordinates) such that $T(\sigma) = T_w$. 

\section{Isobaric Dynamics}
\label{sec_isobaric}

We study here the motion of conduction fronts in isobaric media. Early analyses of this motion were done by Zel'dovich and Pickel'ner \cite{ZP69} (hereafter ZP) and by Penston and Brown \cite{Penston_Brown_1970}. Both pairs of authors derive a condition on the `critical pressure' at which conduction fronts are stationary. In addition, both presented approximation schemes for front motion at non-critical pressures. However, the former treatment was quite \textit{ad hoc} and vague as to the fidelity of the approximation, while the latter was primarily numerical. AMS \cite{Aranson_Meerson_Sasorov_1993} derived a closed-form perturbative expression for the rate of front motion. A later analysis by Elphick, Regev, and Spiegel \cite{Elphick_Regev_Spiegel_1991} (Hereafter ERS) calculated the interactions between neighboring conduction fronts, yielding fundamental insights into the inverse cascade that appears in bistable fluid. This analysis was, however, dependent on a particular cooling function carefully chosen for analytical treatment, and so cannot be quantitatively transferred to other systems. In this section, we apply a systematic asymptotic analysis to derive expressions for the motions of fronts.

\subsection{Isolated fronts}
\label{sec_isobaric_isolated_fronts}

Throughout this section, we enforce that $\partial_t p = 0$. Physically, this corresponds to a wide variety of systems whose container (or surrounding fluid) applies constant pressure. The timescale order in Eq.~\eqref{eq_timescale_ordering} is still used, so $\partial_x p = 0$, and pressure is therefore entirely constant. We derive the conditions under which a conduction front is stationary and, under non-stationary conditions, find a perturbative solution for its motion.

We consider a system consisting of one conduction front separating a semi-infinite cold region from a hot one. We use $T_c$ and $T_h$ to denote the equilibrium temperature in the cold and hot regions respectively. When there is no ambiguity, $T_{c/h}$ serves as shorthand for $T_c$ or $T_h$ to leave some statements general.

Following the treatment of \cite{Elphick_Regev_Spiegel_1991}, we transform to a moving coordinate system by defining $z \doteq \si - J(t)$, where $J$ is a function of time only and $j \doteq dJ/dt$ is a particle flux that we seek to calculate. We seek traveling-wave solutions, for which $T$ is a function of $z$ only. In these coordinates, Eq.~\eqref{eq_dt_langrang_gen} becomes
\begin{equation}
\label{eq_dt_y_fixp}
    -j\heatcap n\frac{dT}{dz} = -\mc L + n\frac{d}{dz}(n \kappa \frac{dT}{dz}) ,
\end{equation}
and, by defining ${q \doteq n \kappa (dT/dz)}$, we can write this as
\begin{equation}
\label{eq_dqdy}
    n\kappa \frac{d q}{dz} + j\heatcap q = \kappa \mc L .
\end{equation}
We will sometimes refer to $q$ as `heat flux' for brevity, even though it differs from the physical heat flux by a (negative) factor.

Assuming that the temperature profile is monotonic, we follow ZP \cite{ZP69} and transform so that the dependent variable is $T$, noting $\partial_z q = (q/n\kappa)(dq/dT)$, and arrive at
\begin{equation}
\label{eq_q_T_full}
    \half \frac{d}{dT}\left(q^2\right) + j\heatcap q = \kappa\mc L .
\end{equation}

Finally, we integrate both sides to obtain
\begin{equation}
\label{eq_q_T_integral_exact}
    \half q^2\Big |_{T_c(p)}^{T_h(p)} + j\heatcap \int_{T_c(p)}^{T_h(p)} dT q(T)  = \kL(p) ,
\end{equation}
where we have defined the `net cooling' function $\Lambda$ as
\begin{equation}
\label{eq_Phi_defn}
    \kL(p) \doteq \int_{T_c(p)}^{T_h(p)} dT \kappa(p,T) \mc L(p, T) .
\end{equation}

We start by considering a stationary front, where $\partial_t T = 0$ and therefore $j = 0$. Far from the front, in the bulk of each phase where the temperature takes one of its equilibrium values, there should be no heat flux\footnote{This is true only if the boundaries can be taken to be arbitrarily deep in the bulk of each phase. If the system size is finite, the $q$ at the boundaries may not be exactly zero, or the temperatures there may not be exactly $T_h, T_c$. This point is discussed further in \ref{sec_boundary_effects}; in general, the corrections are exponentially small in $\delta$.}. Therefore, the boundary terms on the LHS of Eq.~\eqref{eq_q_T_integral_exact} vanish and consistency requires $\kL(p) = 0$. This is the condition derived by ZP \cite{ZP69}, which they term a `phase condition' by analogy to Maxwell's area rule in equilibrium thermodynamics. 

Let $\pcset$ be the set of `critical pressures' such that the condition $\kL(\pcrit)=0$ is satisfied for all $\pcrit \in \pcset$. For non-critical pressures, which yield nonzero $\kL$, Eq.~\eqref{eq_q_T_full} can only be satisfied by nonzero $j$. In other words fronts must move.

In general, there is no closed-form expression for the front velocity at non-critical pressures. In this work, we consider small deviations from critical pressure and use a perturbative procedure to generate asymptotic solutions.

Introducing a small parameter $\eps \ll 1$, let ${p = \pcrit + \tilde p}$, where $\pcrit \in \pcset$ and ${\tilde p \sim \eps \pcrit}$. Similarly, we expand the particle and heat fluxes as ${j = j^\zero + j^\one + ...}$ and ${q = q^\zero + q^\one + ...}$ where larger superscripts denote higher orders in $\eps$. Finally, we expand $\kL$ about $\pcrit$ as ${\kL(p) \approx \kL_p \tilde p}$ where ${\kL_p \doteq d \kL/dp}$. Although the change in $p$ causes the equilibrium temperatures $T_{c/h}$ to shift, we assume that this shift is of order $\epsilon$ or smaller (this assumption is valid in all cases that we will consider).

In the case above where the pressure was critical, i.e. $\eps = 0$, we were able obtain an integral condition Eq.~\eqref{eq_q_T_full} without actually calculating $q(T)$. In this case, we will need to calculate its lowest-order term $q^\zero(T)$. We know that the particle flux vanishes at lowest order, so $j^\zero = 0$. Integrating Eq.~\eqref{eq_q_T_full} and keeping only leading-order terms in $\eps$, we have 
\begin{align}
\label{eq_q0_relation}
    \half \left[q^\zero(T)\right]^2 &= \int_{T_c(\pcrit)}^T dT'\kappa(\pcrit,T') \mc L(\pcrit, T').
\end{align}

Noting that $q$ should vanish infinitely far from the front to all orders in $\eps$, we expand the two remaining terms of Eq.~\eqref{eq_q_T_integral_exact} to find
\begin{align}
\label{eq_q_integral_firstorder}
    \half q^2\Big|_{T_c(p)}^{T_h(p)} + j_1\heatcap \int_{T_c(\pcrit)}^{T_h(\pcrit)} q^\zero(T) dT &\sim \kL_p(\pcrit) \tilde p ,
\end{align}
and for convenience we define the integral over the zeroth-order heat flux as
\begin{equation}
    \Q(p) \doteq \heatcap \int_{T_c(p)}^{T_h(p)} dT\sqrt{2\int_{T_c(p)}^T dT'\kappa(p,T') \mc L(p, T')} .
\end{equation}

Now, using Eq.~\eqref{eq_q0_relation}, we can solve for $j^\one$ to obtain
\begin{equation}
\label{eq_j1_p}
    j^\one = \tilde p \frac{\kL_p(\pcrit)}{\Q(\pcrit)} \hat q ,
\end{equation}
where $\hat q = \pm 1$ indicates the direction of the temperature gradient ($\hat q = 1$ for a $\ch$ front, while $\hat q = -1$ for a ${\hc}$  front). If the pressure is perturbed such that $\kL_p \tilde p > 1$, then hot material will 'condense' into cold material across the front as it propagates into the hot region; if $\kL_p \tilde p < 1$, then cold material evaporates into hot. Since $\kL$ is just a weighted integral of the cooling function,  this behavior is expected; positive $\kL$ indicates that cooling processes dominate over heating ones. Typically, $\kL_p > 0$ because the cooling rate tends to increase with density more quickly than the heating rate does.


\subsection{Interacting fronts}
\label{sec_isobaric_interacting_fronts}

A realistic system is likely to contain many regions of hot and cold fluid, rather than the pair of semi-infinite regions considered in the previous section. Each of these regions can be viewed as a pair of fronts. In this section, we consider the interaction between these fronts. To see how this interaction arises, consider the boundary conditions of Eq.~\eqref{eq_q_T_integral_exact}, which assume that the temperature profile comes arbitrarily close to the equilibrium temperatures $T_{c/h}$ in the cold and hot regions. Far from the front, as we will demonstrate shortly, the approach to equilibrium temperature is exponential. Therefore, at any finite distance $\sigma$ from the front, a finite difference will remain between $T(\sigma)$ and $T_{c/h}$. If at least one of the regions bordering a front is of finite size, then Eq.~\eqref{eq_q_T_integral_exact} no longer holds exactly.

For concreteness, we consider a hot `bubble' embedded in cold fluid, or, equivalently, a pair of fronts arranged as $\ch$ on the left and $\hc$ on the right. The results of this section will also apply to the opposite situation, a cold cloud in a hot background. In Lagrangian coordinates, the distance between the front cores is $\Scloud$ at $t=0$, and we require the bubble to be large compared to the fronts ($\Scloud \gg n_0l_0$). The cold regions to the left and right of the bubble can be taken to be semi-infinite. We center the coordinate system so that the point $\sigma = 0$ lies at the midpoint of the hot region. The core of the $\ch$ front is therefore given by ${T(\sigma=-\half\Scloud) = T_w}$. Throughout this section, all functions will be evaluated at $t=0$, and so the $t$ argument is dropped for concision.

We define a temperature $h$ such that the temperature at the middle of the hot region is $T(0) = T_h - h$. Importantly, $h\ll T_h$ because the midpoint is far from both fronts. Putting aside for a moment the problem of calculating $h$, we compute the motion of each front in terms of $h$. We seek a traveling-wave solution for the $\ch$ front; since the front motion depends on the size of the bubble, $j$ will no longer be constant. Symmetry allows us to consider just the left-hand front; the other front moves with the same speed in the opposite direction.  Starting from Eq.~\eqref{eq_q_T_full} and integrating, we have
\begin{equation}
\label{eq_integral_h}
    \half q^2\Big|_{T_c}^{T_h-h} + j\heatcap \int_{T_c}^{T_h-h}dT q(T) = \int_{T_c}^{T_h-h} dT \kappa(T)\mc L(T) ,
\end{equation}
where notation of the pressure dependence is dropped for concision.

For simplicity, we take the pressure to be equal to some critical pressure $\pcrit \in \mc P$, meaning that $\kL(\pcrit) = 0$. All front motion, therefore, results from interactions between fronts. To write an analog to Eq.~\eqref{eq_q_integral_firstorder}, we need to apply appropriate boundary conditions. Far to the left, in the cold phase, $q\rightarrow 0$ at all orders. At the middle of the bubble, we know by symmetry that ${q(\sigma=0)=0 ~\forall t}$, or equivalently ${q(T = T_h-h) = 0}$, where $q$ is the \textit{exact} heat flux. Solving Eq.~\eqref{eq_integral_h} for $j$, we find
\begin{equation}
    \label{eq_j_h_exact}
    j = \frac{-\int_0^{h} d\theta \kappa(T_h - \theta) \mc L(T_h - \theta)}{\heatcap \int_{T_c}^{T_h-h}dT q(T)}
\end{equation}

Let us define a small parameter $\dtail \doteq \exp\{-m\Scloud/l_0\}$, where $m$ is an unknown positive constant, such that ${h \sim \mc O(\dtail)}$ (which will be validated later). Now we expand in $\dtail$, writing ${q = q^\zero + q^\one + ...}$ and ${j = j^\zero + j^\one + ...}$; at first order, this is equivalent to expanding in $h$. We also expand $h$ as a series in $\dtail$ as $h = h^\zero + h^\one + ... ~$. The unperturbed heat flux $q^\zero$ is still given by Eq.~\eqref{eq_q0_relation}. `Unperturbed' quantities mean those that would apply if we did not impose the $q(0) = 0$ boundary condition but rather allowed the solution to extend to infinity in both directions as in \S~\ref{sec_isobaric_isolated_fronts}.

To expand the numerator in $h$, we define
\begin{equation}
    \lambda_h \doteq \left(\frac{\partial}{\partial T}\kappa \mc L\right)\Big|_{T_h}
\end{equation}
and, since $\mc L(T_h) = 0$ by definition, we arrive at a leading-order expression for $j$:
\begin{equation}
\label{eq_j1_h}
    j^\one = \frac{\left(h^\zero\right)^2\lambda_h }{2 \Q } .
\end{equation}

The task is now to calculate $h$ in terms of known quantities. Let $T^\zero(z)$ be the temperature profile corresponding to $q^\zero$ and $j^\zero$. Then the leading-order term in $h$ is given by $h^\zero = T^\zero(0)$. 

To calculate $h^\zero$, we need to find information about the temperature profile $T^\zero(z)$. We will accomplish this through asymptotic matching. The structure of the problem suggests slowly varying `outer' regions in the bulk of the cold and hot phases and a rapidly varying `inner' region near the core of the front. We will first solve for the leading-order outer solution in the hot phase, up to an unknown constant. Next, we will solve for the leading-order inner solution. Identifying an intermediate region in which both solutions can be expanded, we will match the solutions in this region to determine the unknown constant. Finally, since the outer solution's region of validity contains $\sigma=0$, we can determine $h^\zero$.

Let $\tau \doteq T_h - T$. In the hot outer region, $\tau \sim \mc O(\dtail)$. To leading order, Eq.~\eqref{eq_q_T_full} becomes
\begin{equation}
    \half \frac{d}{d\tau} \left(q^\zero\right)^2 = \lambda_h \tau ,
\end{equation}
which we can solve to find
\begin{equation}
    q^\zero(\tau) = \sqrt{\lambda_h\tau^2 + a} .
\end{equation}
Noting that the unperturbed solution should satisfy ${\lim_{\tau \rightarrow 0} q^\zero(\tau) = 0}$, we see that $a=0$. Now we solve for $T^\zero$. In the outer region, ${q^\zero = -n_h\kappa(T_h)\partial_z\tau}$ (note $n_h = p/T_h$). The leading-order solution for $\tau(z)$ is therefore
\begin{equation}
\label{eq_tau_z_exp}
    \tau(z) \sim \tau_0 \exp\left\{-\frac{\sqrt{\lambda_h}}{n_h\kappa(T_h)}z\right\} ,
\end{equation}
where $\tau_0$ is an unknown constant. We define ${k_h \doteq \sqrt{\lambda_h}/n_h\kappa(T_h)}$ for convenience\footnote{Note this is equivalent to $k_h = (T_h/p)\sqrt{(\partial\mc L/\partial T)|_{T_h} / \kappa(T_h)}$. Therefore, ${k_h\Scloud \sim \mc O(L/l_0)}$ where $L$ is the size of the cloud in ordinary spatial (Eulerian) coordinates.}. Inverting Eq.~\eqref{eq_tau_z_exp}, we find
\begin{equation}
\label{eq_z_tau_outer}
    z(\tau) \sim -\frac{1}{k_h}\ln(\tau) + \frac{1}{k_h}\ln(\tau_0) .
\end{equation}

We proceed to the inner region, which contains the front core and the rapidly varying part of the solution. Here, we find an integral equation, which applies even in the outer region but is intractable. The final task is then to match the inner solution to the closed-form outer solution in Eq.~\eqref{eq_tau_z_exp}. Keeping leading-order terms in Eq.~\eqref{eq_q_T_full} and rearranging gives
\begin{equation}
    dz \sim dT\frac{n(T)\kappa(T)}{\sqrt{2\int_{T_c}^T dT' \kappa(T')\mc L(T')}}.
\end{equation}

Using the condition $z(T_w) = -\half \Scloud$, we find that $z(\tau)$ is, to leading order,
\begin{equation}
\label{eq_z_tau_inner}
    z(\tau) \sim -\half \Scloud + \int_{\tau}^{T_h-T_w} \frac{n(T_h-\tau')\kappa(T_h-\tau')d\tau' }{\sqrt{-2\int_0^{\tau'} d\theta \kappa(T_h -\theta)\mc L(T_h-\theta)}} .
\end{equation}

Note that $z$ has a logarithmic divergence as $\tau \rightarrow 0$ because the denominator is approximately linear in $\tau'$ in this limit. Physically this is appropriate because, in order for $T$ to be close to $T_h$, we must be very far from the front, meaning $z$ is large. Beyond the logarithmic term, the expansion of $z$ in $\tau$ consists of a constant term followed by positive powers of $\tau$.

To determine $\tau_0$, we now match Eq.~\eqref{eq_z_tau_inner} and Eq.~\eqref{eq_z_tau_outer} to leading order in $\dtail$, which entails matching the constant (with respect to $\tau$) terms and discarding the terms in Eq.~\eqref{eq_z_tau_inner} that vanish as $\tau \rightarrow 0$. The logarithmically divergent part of Eq.~\eqref{eq_z_tau_inner} should equal the logarithmic term of Eq.~\eqref{eq_z_tau_outer}.
Equating the remaining terms in both equations to leading order, we arrive at
\begin{equation}
    \ln(\tau_0) = -\half k_h \Scloud + \ln(T_h-T_w) + \Theta_h
\end{equation}
where
\begin{equation}
\label{eq_Theta_def}
    \Theta_h \doteq \int_0^{T_h-T_w} d\tau' \left[\frac{k_h n(T_h-\tau')\kappa(T_h-\tau')}{\sqrt{-2\int_0^{\tau'} d\theta \kappa(T_h -\theta)\mc L(T_h-\theta)}} - \frac{1}{\tau'}\right] .
\end{equation}

In Eq.~\eqref{eq_Theta_def}, the lower bound of the integral is taken to zero because we are only interested in leading-order behavior and so can ignore all higher powers in $\tau$. It is straightforward to show that $\Theta$ is finite when $\mc L$ has the properties described in \S\ref{sec_cooling_curves}. Using Eq.~\eqref{eq_tau_z_exp} to determine $h$ to leading order, we find
\begin{equation}
\label{eq_h0}
    h^\zero = (T_h - T_w) e^{\Theta_h - \half k_h \Scloud} ,
\end{equation}
validating the form that we assumed for $\dtail$. Combining Eq.~\eqref{eq_h0} and Eq.~\eqref{eq_j1_h}, we have an explicit expression for $j^\one$. It is straightforward to generalize to a cold cloud embedded in a hot background. We can define quantities $\lambda_c, k_c, \Theta_c$ analogously to their hot counterparts; explicit definitions are given in Appendix~\ref{sec_appendix_definitions}. 

For generality, let the subscript $g$ ($=h$ or $c$) denote either the hot or the cold phase. We consider a cloud of phase $g$ and size $\Scloud$ immersed in much larger regions of the other phase. The fronts surrounding the cloud will move toward each other with a flux described by
\begin{equation}
\label{eq_j_Sigma_general}
    j^\one = \Gamma_g e^{-k_g\Scloud} \hat q ,
\end{equation}
where $\hat q$ now points toward the interior of the cloud ($\hat q = 1$ for the left-hand front and $\hat q = -1$ for the other) and
\begin{equation}
\label{eq_Gamma_def}
    \Gamma_g \doteq \frac{\lambda_g |T_g - T_w|^2 e^{2\Theta_g}}{2\Q} .
\end{equation}

Assuming all other conditions remain unchanged, this motion causes the cloud size to decrease at the rate ${d\Scloud/dt = -2|j|}$. The cloud size over time is described by
\begin{equation}
\label{eq_Sigma_t}
    \Scloud(t) = \frac{1}{k_g} \ln\left(e^{k_g\Scloud(0)} -  2\Gamma_g k_g t \right) .
\end{equation}

\subsection{Boundary effects}
\label{sec_boundary_effects}

The focus of this work is systems in a finite `box.' When front interactions are included, boundary conditions imposed at the ends of the box can affect front motion. To avoid introducing new parameters to the problem, we impose Neumann boundary conditions. For a system of length $L_0$, we require that ${\partial_x T = 0}$ at $x=0$ and $x=L_0$. This is equivalent to requiring $q=0$ at both points. Note that this precisely the condition that we imposed at the midpoint of each region in \S\ref{sec_isobaric_interacting_fronts}. Therefore, a region of size $\Scloud$ abutting one of the boundaries interacts with its one neighbor as if it were an ordinary region of size $2\Scloud$ generated by `reflecting' the real region across the boundary.

For the rest of this work, unless stated otherwise, statements about the size of each region should be taken as applying to this fictitious doubled region when referring to the regions on the boundaries. Indeed, the entire system can be reflected across each boundary and so viewed as periodic with period $2L_0$.

A real system will generally not satisfy this boundary condition exactly; any realistic container can be expected to admit some heat flux through the wall and so fail to satisfy $\partial_x T = 0$. However, for most systems that we will consider, the boundary will not play a major role and so this approximate boundary condition is a reasonable one.

\subsection{Cloudy medium}
\label{sec_isobaric_cloudy_medium}

With Eq.~\eqref{eq_j_Sigma_general}, we are equipped to describe the evolution of a system containing many hot and cold regions. The lifetime 
\begin{equation}
\label{eq_cloud_lifetime}
    t_\mathrm{life} = (e^{k_\ph\Scloud(0)} - 1)/2\Gamma_\ph k_\ph
\end{equation} of a single cloud increases exponentially with the cloud size for large clouds. However, any cloud of finite size will eventually contract. This means that a two-phase system of this type will typically not be in a true steady state, even at critical pressure.  This fact was explored by ERS \cite{Elphick_Regev_Spiegel_1991} (see also \cite{Elphick_Regev_Shaviv_1992}). A true steady state can exist when the medium is uniform (all hot or cold phase). If the system is large enough that the boundary can be ignored, then a configuration with one front separating two semi-infinite regions is also a steady state.

As a more general scenario, let us consider a `cloudy' medium consisting of $N$ alternating hot and cold regions indexed as ${i = 1,2, ... ~N }$ and let $\Scloud_i$ be the size of front $i$. Hereafter, we denote by $\ph$ the phase of region $i$ and by $\php$ the other phase. For each front, the finite-size effects from the cloud on each side simply add (at the order to which we are working), so the velocity $j_i$ of the front on the left side of cloud $i$ (where $i>0$) is, to leading order,
\begin{equation}
\label{eq_j1_h_twosides}
    j_i^\one = -\Gamma_{\php}e^{-k_{\php}\Scloud_{i-1}} + \Gamma_\ph e^{-k_\ph \Scloud_i} .
\end{equation}

The time evolution of region $i$ (where $i \in (1, N)$) is then described by
\begin{equation}
\label{eq_dSigma_i_dt}
    \frac{d\Scloud_i}{dt} = -2\Gamma_\ph e^{-k_\ph\Scloud_i} + \Gamma_{\php} \left[e^{k_{\php}\Scloud_{i-1}} + e^{k_{\php}\Scloud_{i+1}}\right] .
\end{equation}

Heuristically, smaller regions shrink more quickly. The calculation here provides additional nuance that does not appear in the model of ERS; the most rapidly shrinking region is not necessarily the one with the least material (minimal $\Scloud_i$) but rather the one for which the weighted quantity ${\hat\Scloud_i \doteq k_\ph\Scloud_i - \ln\Gamma_\ph}$ is minimal\footnote{Even this statement can be further qualified. For a complex cloudy medium where many clouds are of comparable size, the evaporation rate of each cloud is influenced by its neighbors, so statements about which clouds shrink most quickly and vanish first depend on detailed knowledge of the system configuration.}. 

Over time, this cloudy medium evolves as described by ERS, except for the above caveat that `cloud size' should be evaluated using $\hat \Scloud$. The smallest clouds evaporate first, donating their material to their neighbors. When a cloud evaporates completely, its neighbors merge, becoming more robust against shrinking due to their larger size. As the surviving clouds become larger, the process slows exponentially, but still completes in finite time. Steady state is reached when only one uniform region remains (or, discounting system boundaries, two regions and one front).

This evolution leads to an inverse cascade, destroying small-scale structure and creating large-scale coherent structure in a process similar to Ostwald ripening. Several works have indicated that large clouds tend to `shatter' due to various instabilities, providing a limitation on the inverse cascade \cite{Jennings_Li_2021}. This shattering typically appears in multidimensional simulations with self gravity, so it is no surprise that our results in this reduced model do not include such a process. 

A complex steady state is possible. This requires a periodic arrangement of hot and cold regions such that $\hat\Scloud_i$ is equal for all $i$, as seen in Eq.~\eqref{eq_dSigma_i_dt}. However, this arrangement is unstable. Perturbing the location of a front to make one region slightly smaller and one slightly larger will cause collapse of the small region and subsequently of the whole arrangement.

\section{Isochoric Dynamics}
\label{sec_isochoric}

Up to this point, we have considered isobaric systems, in which the surroundings apply constant pressure to the system. The use of Lagrangian coordinates permitted us to ignore many details of the fluid flow; however, we know that the system must expand and contract to maintain hydrostatic equilibrium between hot regions, cold regions, and the surroundings. We now consider an isochoric system, whose volume is fixed but whose pressure is allowed to vary. Because of the ordering in Eq.~\eqref{eq_timescale_ordering}, pressure remains uniform throughout the system; $p$ is a function of $t$ only. Individual regions continue to expand and contract in response to temperature variations but if, for example, evaporation causes the fraction of fluid in the hot phase to increase, pressure must increase everywhere. 

\subsection{Evolution of an isolated front}
\label{sec_isochoric_isolated_front}

We begin by considering a system of two regions separated by a conduction front. Boundary effects are ignored for the moment. One region is in the hot phase and the other is in the cold phase. The condition under which this front is stationary (Eq.~\eqref{eq_q_T_full}) is the same as in the isobaric case. When this condition is not satisfied, the front will move, but now, crucially, this motion will cause $p$ to change. We denote the temperature and density in the region on the left  by $T_1, n_1$ and in the region on the right by $T_2, n_2$. The time evolution of the pressure can be written as follows:
\begin{align}
\label{eq_dpdt_eos}
    \frac{dp}{dt} = n_i \frac{dT_i}{dt} + T_i \frac{dn_i}{dt}
\end{align}
for $i = 1, 2$. This relation applies in both regions.

As the conduction front propagates, the fractional expansion of the bulk fluid on each side in one cooling time $t_\mathrm{cl}$ is on the order of $\delta$. Therefore, the timescale on which adiabatic expansion heats and cools bulk fluid by the same amount as external processes is $t_a = t_\mathrm{cl}/\delta$. Then as a consequence of Eq.~\eqref{eq_timescale_ordering}, $t_a \gg t_\mathrm{cl}$, meaning that external processes are fast enough to keep the bulk of each phase close to its equilibrium temperature during front motion. Denoting by $T_i\eq$ the equilibrium temperature\footnote{There is a notational subtlety here: $T_i$ is the fluid temperature in regions $i=1,2$, while $T_i\eq(p)$ is the function describing the locus of zeros of the cooling curve $\mc L (p, T) = 0$. The subscript indicates (in the bistable region) which of the solutions the fluid temperature matches; in practice, this distinction is unimportant and $T_i = T_i\eq$.} in the region under consideration, the time evolution of $T_i$ is determined to leading order in $\delta$ by
\begin{equation}
\label{eq_dTdt_Teq}
    \frac{dT_i}{dt} = \frac{dT_i\eq}{dt} = \frac{dT_i\eq}{dp} \frac{dp}{dt} .
\end{equation}

We define for future use
\begin{equation}
\label{eq_lambda_defn}
    \symbdTdp_i \doteq \frac{p}{T_i}\frac{\mc L_p}{\mc L_T} = \frac{p}{T_i}\frac{dT_i\eq}{dp} ,
\end{equation}
where again $T_i\eq$ represents the equilibrium temperature in region $i$.

To use Eq.~\eqref{eq_dpdt_eos}, we need to know how the density evolves. This happens in two ways: expansion or contraction of each region; and evaporation (or condensation) of material from one phase into the other across the conduction front. Let $r_1(t) \in [0, L_0]$ denote the location (in real space) of the conduction front at time $t$, and $r_2 = L_0 - r_1$. As before, $j(p)$ denotes the particle flux across the front. The density then evolves according to
\begin{equation}
\begin{split}
\label{eq_dn12_dt}
    \frac{dn_1}{dt} &= -\frac{n_1}{r_1}\frac{dr_1}{dt} + \frac{j}{r_1} ,
    \\
    \frac{dn_2}{dt} &= -\frac{n_2}{r_2}\frac{dr_2}{dt} - \frac{j}{r_2} .
\end{split}
\end{equation}

It is more convenient to keep track of the region sizes using areal densities $\Scloud_1$ and $\Scloud_2$, defined as $\Scloud_i \doteq n_i r_i$. Substituting Eq.~\eqref{eq_dn12_dt} and Eq.~\eqref{eq_dTdt_Teq} into Eq.~\eqref{eq_dpdt_eos}, we obtain
\begin{equation}    
\label{eq_dpdt_chi}
    \frac{dp}{dt} = j \chi_{1;2}(p) ,
\end{equation}
where
\begin{equation}
\label{eq_chi_defn}
    \chi_{i;j}(p) \doteq \frac{(T_i\eq - T_j\eq) p}{(1 - \symbdTdp_i)T_i\eq \Scloud_i + (1 - \symbdTdp_j)T_j\eq\Scloud_j} .
\end{equation}

Explicit models for the cooling function will be discussed in \S~\ref{sec_numerics_cooling_models}. In brief, we note here that, for a typical cooling curve, $\symbdTdp < 0$ for the stable phases\footnote{Furthermore, it is generally the case that $\symbdTdp$ is of order unity. This validates the assumption, made in deriving Eq.~\eqref{eq_j1_p}, that the variation in $T_{c/h}$ due to small deviations from critical pressure is small. Such variations in equilibrium temperatures are of order $\eps \symbdTdp$, and since they would only appear multiplied by the ${\kL_p \tilde p \sim \mc O(\eps)}$ term, they do not contribute at leading order in Eq.~\eqref{eq_j1_p}.}. For such curves, the denominator remains positive for all values of $p$.

If we now wish to calculate the motion of the front, the $\partial_t p$ term in Eq.~\eqref{eq_dt_langrang_gen} is no longer identically zero. In the next section, we show that this effect can safely be neglected.

Suppose that the pressure is near a critical pressure so that $p = \pcrit + \tilde p$ and $\tilde p \sim \eps \pcrit$. Naturally, $\pcrit$ is a constant. Using Eq.~\eqref{eq_j1_p}, the rate of change of pressure, to first order in $\eps$ and zeroth order in $\delta$, is
\begin{equation}
\label{eq_dpdt_firstorder_isolated}
    \frac{d\tilde p }{dt} \sim -\tilde p \frac{\kL_p(\pcrit) \chi_{h;c}(\pcrit) }{\Q(\pcrit)}
\end{equation}
(the choice to write $\chi_{h; c}$ rather than $\chi_{c; h}$ accounts for the sign of $\hat q$ and makes the result insensitive to the placement of hot and cold regions).

So far, we have adopted a maximal ordering of $\eps$ and $\delta$. One might then wonder whether Eq.~\eqref{eq_dpdt_firstorder_isolated} fails to capture the dominant behavior when $\delta \gg \eps$ (a condition that is guaranteed to be met eventually as $L_0$ is held constant while $p \rightarrow \pcrit$. Note, however, that $j(\pcrit) = 0$ is an exact result. Nonzero $j$ only appears when $\eps \neq 0$, so there are no $\mc O(\eps^0 \delta^k)$ terms (for any $k$) missing from Eq.~\eqref{eq_dpdt_firstorder_isolated}.

In the vicinity of a critical pressure, $\tilde p(t)$ is exponential in time. The terms multiplying $\tilde p$ on the RHS of Eq.~\eqref{eq_dpdt_firstorder_isolated} give the growth rate. Assuming that $\symbdTdp_{c/h} < 1$, the sign of the growth rate is determined by $\kL_p$. For $\kL_p > 0$, the pressure decays toward its critical value; for $\kL_p < 0$, the pressure perturbation grows exponentially. This represents a secondary instability--the primary being the classical thermal instability--that appears for class of cooling functions whose $\mc P$ contains a pressure $\pcrit \in \mc P : \Lambda_p(\pcrit) < 0$. This instability has not previously been discussed in literature on the isochoric thermal instability, and it may be of physical significance in experiments because it drives a sudden change in pressure. It is, however, not clear that the primary thermal instability will cause a general initial perturbation to evolve into a state susceptible to the secondary instability; this question requires further study.

\subsection{Pressure variation}
\label{sec_pressure_variation}

In the last section, we saw that the motion of a single front in an isochoric system causes the pressure to change, though this change does not feed back to influence the front speed at leading order. Here, we show that pressure variation, when driven by some external source, causes conduction fronts to move.

We consider a system initially at critical pressure $\pcrit$ whose pressure is varying at a rate $\nu$ such that $\partial_t p = \nu \pcrit$. This variation could be accomplished by uniform, slow compression of the system, for example. Starting with Eq.~\eqref{eq_dt_langrang_gen} and changing variables to work with $q(T)$, we have
\begin{equation}
    \half \frac{d}{dT}\paren{q^2} + jq = - \kappa \nu \pcrit .
\end{equation}

If we are considering a single isolated front, with boundary effects unimportant, then analysis proceeds as in \S~\ref{sec_isobaric_isolated_fronts}, the only difference being that the inhomogeneous driving term on the RHS corresponds to pressure variation instead of external heating and cooling. Much as the earlier analysis relied on $\eps \ll 1$, applying the same treatment here requires $\nu t_\mathrm{cl} \ll 1$. Assuming this holds, the analogous result to Eq.~\eqref{eq_j1_p} is
\begin{equation}
\label{eq_j1_nu}
    j^\one = -\frac{\nu \pcrit K(\pcrit)}{\Q(\pcrit)} \hat q ,
\end{equation}
where we have defined
\begin{equation}
\label{eq_K_def}
    K(p) = \int_{T_c}^{T_h} dT \kappa(p, T) ,
\end{equation}
and as a reminder, $\hat q = \pm 1$, with its sign positive for a $\ch$ front and negative for a $\hc$ front. Eq.~\eqref{eq_j1_nu} therefore has a surprising consequence: if pressure is increasing ($\nu > 0$), conduction fronts move from hot to cold. This further increases the pressure. 

\section{Evolution and Equilibrium}
\label{sec_evolution}

\subsection{Consistency of assumptions}

We are now able to describe the evolution of an ensemble of fronts (i.e. a cloudy medium) in a fixed volume with pressure treated as a dynamical variable. Three processes can drive front motion:
\begin{enumerate}
    \item Non-critical pressure ($\kL \neq 0$),
    \item Finite cloud size ($k_\ph\Scloud_i < \infty$),
    \item Time-varying pressure ($\partial_t p \neq 0$).
\end{enumerate}

For the asymptotic analysis in this work, several parameters must be small. The condition that $\delta \ll 1$ (conduction fronts are much smaller than the full system) is satisfied by an appropriate choice of fixed parameters. The following conditions, which depend on dynamical quantities, must be met at all times:
\begin{enumerate}
    \item Small pressure perturbations ($\eps \ll 1$),
    \item Large clouds ($k_\ph \Scloud_i \gg 1$ and thus $\dtail \ll 1$),
    \item Slow pressure variation ($\nu t_\mathrm{cl} \ll 1$).
\end{enumerate}

As an initial condition, we assume that the pressure is in the vicinity of $\pcrit$ so that Condition 1 holds, at least initially. At future times, pressure evolves self-consistently but without external forcing (unlike in \S\ref{sec_pressure_variation}). If $\kL_p < 0$ then, as found in \S\ref{sec_isochoric_isolated_front}, pressure perturbations are unstable and Condition 1 will rapidly be violated. In that case, the system will quickly evolve toward the neighborhood of a stable critical pressure $p^{**} \in \mc P$ and analysis can proceed from there. From now on, we suppose that $\kL_p(\pcrit) > 0$. In this case, perturbations from critical pressure are suppressed and so we can expect $\eps$ to remain small.

Because no external forces are driving pressure variation, the only contribution to $\partial_t p$ comes from the combined motion of all fronts. 
We now show that this permits us to neglect $\partial_t p$ in analyzing the system dynamics.

The leading-order contribution to $j$ for each front scales as the larger of $\eps$ and $\dtail^2$. Furthermore, $\partial_t p$ is suppressed relative to $j$ by an $\mc O(\delta)$ factor. Thus, $\nu t_\mathrm{cl} \lesssim  \mc O \left(\max (N\delta \eps, N\delta \dtail^2) \right)$, where $N$ is the number of fronts. We suppose that $N$ is not asymptotically large, i.e. $N\delta \ll 1$. In this case, not only is Condition 3 satisfied, but $\nu t_\mathrm{cl}$ is of sub-leading order, meaning that the rate of pressure variation due to front motion can be neglected in determining the motion of other fronts.

The final assumption, Condition 2, is guaranteed to be violated eventually for nearly all initial conditions. Suppose that, initially, all clouds are large enough that Condition 2 holds. As we saw in \S\ref{sec_isobaric_cloudy_medium}, the smallest clouds tend to shrink and eventually to disappear. As a cloud becomes small, the shrinking process accelerates. At some point in this process, when the size of some cloud comes to be on the order of the front width, Condition 2 will be violated. Although Eq.~\eqref{eq_dSigma_i_dt} ceases to apply at that point, the analysis in \S\ref{sec_isobaric_interacting_fronts} makes clear qualitatively that the small cloud will rapidly finish evaporating. Note, however, that the cloud mass is still less than the total system mass by an $\mc O(\delta)$ factor, so the resulting pressure change will not violate Condition 3.


\subsection{Evolution of a cloudy medium}
\label{sec_isochoric_cloud_medium}

Combining the above results, we can write a complete system of equations describing the evolution of a `cloudy' system consisting of many hot and cold regions. The setup and labeling convention are as described in \S\ref{sec_isobaric_cloudy_medium}. Transplanting results is straightforward, excepting the breakdown of our assumptions when any cloud becomes too small. To isolate this case, we define a cutoff cloud size $\hat \Scloud_\mathrm{crit}$ and treat as a special case the state where any cloud has size $\hat \Scloud_i<\hat \Scloud_\mathrm{crit}$.

Separate treatment of this special case is unnecessary if the scale separation between the system size $L_0$ and the front width $l_0$ is large enough that ${\delta \ll \eps, \eta^2}$. However, much of the interesting physics resulting from the interplay of pressure-driven and front interaction-driven motion is most relevant when $\delta$ is not too small.

The following equations apply when $\hat \Scloud_i \geq \hat \Scloud_\mathrm{crit}~\forall i$:
\begin{align}
\label{eq_time_evol_full_sig}
    \frac{d\Scloud_i}{dt} &= -2\Gamma_\ph e^{-k_\ph\Scloud_i} + \Gamma_{\php} \left[e^{-k_{\php}\Scloud_{i-1}} + e^{-k_{\php}\Scloud_{i+1}}\right] ,
    \\
\label{eq_time_evol_full_p}
    \frac{dp}{dt} &= \chi_{h;c} \sum_i \delta_{g,h} \frac{d \Scloud_i}{dt},
\end{align}
with the caveat that both $d\Scloud_i/dt = 0$ and $dp/dt = 0$ if there is only one remaining region.

The symbol $\delta_{g,h}$ is a Kronecker delta selecting only hot regions. Since the total particle number is conserved, the increase in areal density in the hot regions is equal and opposite to that in the cold regions. The first line is identical to Eq.~\eqref{eq_dSigma_i_dt} and the second line is an extension of Eq.~\eqref{eq_dpdt_chi}.

If for any region $i$, $\hat \Scloud_i  < \hat\Scloud_\mathrm{crit}$, then we treat the final evaporation of this region by the following prescription. First, define
\begin{equation}
\label{eq_Deltap_defn}
    \Delta p \doteq -\Scloud_i \chi_{\ph;\php}(p).
\end{equation}

Region $i$ should now be removed from consideration and (if it was not on one end of the system) the two regions bordering it should be combined. All regions should be relabeled such that indices are sequential and $N$ represents the new number of regions -- generally there are two fewer than previously. Next, we update the size of all other regions in the following way:
\begin{equation}
    \Scloud_\ell \rightarrow \Scloud_\ell + \frac{(2\delta_{g_\ell,h} - 1)K(p)}{\Q(p)}\Delta p  ,
\end{equation}
where the first factor in the numerator ensures that the sign is correct for hot and cold regions.

Finally, the pressure is updated:
\begin{equation}
    p \rightarrow p + \Delta p .
\end{equation}

This procedure is approximate; we resorted to it because the final evaporation of region $i$ was no longer described by our asymptotic procedure. Physically, we have approximated the sudden vanishing of region $i$ by an instantaneous jump in pressure.

In \S\ref{sec_numerics}, we examine numerically the behavior of solutions to this system of equations. However, for a simplified set of initial conditions, an analytical treatment is possible; we present this in the following section.

\subsection{Equilibrium of an isolated cloud}
\label{sec_evol_isolated_cloud}

We now consider a cloud of cold fluid, embedded a box of hot fluid of finite size, at a critical pressure $\pcrit$ that is stable against pressure perturbations ($\kL_p > 0$). In the isobaric case, we saw in \S~\ref{sec_isobaric_interacting_fronts} that this cloud will eventually evaporate. In the isochoric case, Eq.~\eqref{eq_dpdt_firstorder_isolated} provides a mechanism to stop this evaporation. As the cloud shrinks (Eq.~\eqref{eq_dSigma_i_dt}), the system pressure will rise. As $\kL > 0$, the cold phase becomes favored according to Eq.~\eqref{eq_j1_p}. At some equilibrium pressure $p'$ and equilibrium cloud size $\Scloud'$, these effects will balance and contraction will stop, unless the cloud has already evaporated completely.

Equilibrium is reached when the fronts delimiting the cloud are stationary. Motion of these fronts is driven by noncritical pressure and by interfront interactions. Let $\Scloud_0$ be the size of the full system and let the cloud be centered such that the hot regions on either side have size $(\Scloud_0-\Scloud')/2$. The cloud must be in steady state, meaning that the LHS of Eq.~\eqref{eq_time_evol_full_sig} is zero. This condition reads
\begin{equation}
\label{eq_equilib_pprime_Sigmaprime}
    \frac{\kL(p')}{\Q(p')} = \Gamma_c e^{-k_c\Scloud'} - \Gamma_h e^{-k_h(\Scloud_0-\Scloud')}.
\end{equation}
Of course, a similar expression could be written for a hot cloud embedded in cold background by replacing the appropriate subscripts.

As long as the cloud is not so small that $\dtail \sim \mc O(1)$, the perturbation introduced by interfront interactions is small, so the equilibrium pressure should be close to the critical pressure: $p' - \pcrit \sim \mc O(\dtail^2)$. Then each function of pressure can be expanded about $\pcrit$ and we find that, to leading order, Eq.~\eqref{eq_equilib_pprime_Sigmaprime} reduces to
\begin{widetext}
\begin{equation}
\label{eq_equilib_pprime_Sigmaprime_expanded}
    p'(\Scloud') \sim \pcrit + \frac{\lambda_c (T_w-T_c)^2 e^{2\Theta_c} e^{-k_c\Scloud'} -  \lambda_h(T_h - T_w)^2 e^{2\Theta_h} e^{-k_h(\Scloud_0-\Scloud')}}{2\kL_p(\pcrit)} .
\end{equation}
\end{widetext}

If the cloud is much smaller than the full system, the second term in the numerator is generally very small (by a factor of ${\mc O\paren{\exp\left\{-k_h\Scloud_0 +(k_h+k_c)\Scloud'\right\}}}$ compared to the first and could be dropped to yield an even simpler expression.

This extends the results of AMS \cite{Aranson_Meerson_Sasorov_1993}, who assert that pressure approaches $\pcrit$ and remains at that value for long times. They insightfully note that interfront interactions will eventually cause clouds to collapse on long timescales and predict that this will typically produce a state with one front. In this section, we have provided a quantitative description of this intuition. We have further shown that the pressure does not strictly speaking approach $\pcrit$, but rather approaches the value in Eq.~\eqref{eq_equilib_pprime_Sigmaprime_expanded} that establishes equilibrium between pressure-driven and interfront interaction-driven motion.


\section{Numerical Solutions}
\label{sec_numerics}

The evolution equations derived in \S\ref{sec_isobaric_cloudy_medium} contain rich structure but are relatively opaque to analytical treatment. In this section, we explore their consequences numerically. First, we discuss explicit models for $\mc L$. Next, we consider the evolution of an isolated cloud and compare to the analytical theory developed in \S\ref{sec_evol_isolated_cloud}. Finally, we simulate the evolution of a cloudy medium.

\subsection{Cooling models}
\label{sec_numerics_cooling_models}

\begin{widetext}    
As a model for the cooling function, we adopt the formula\footnote{We have converted the units and notation to match those used elsewhere in this work.} used by Inoue et al. \cite{Inoue_Inutsuka_Koyama_2006}, viz.
\begin{equation}
\label{eq_mcL_IIK}
    \mc L(p, T) = -1.2\times10^{-14} \frac{p}{T} + 4.6\times 10^{-9} \frac{p^2}{T^2} \exp\left\{\frac{-10.2}{T + 0.129}\right\} + 4.9\times 10^{-15} \frac{p^2}{T^2}\exp\left\{\frac{-7.93\times 10^{-3}}{T}\right\} ,
\end{equation}
where $T$ and $p$ are measured in $\mathrm{eV}$ and $\mathrm{eV}~\mathrm{cm}^{-3}$ respectively, and $\mc L$ is measured in $\mathrm{eV}~\mathrm{cm}^{-3}~\mathrm{s}^{-1}$.
\end{widetext}

\begin{figure}
    \centering
    \includegraphics[width=\columnwidth]{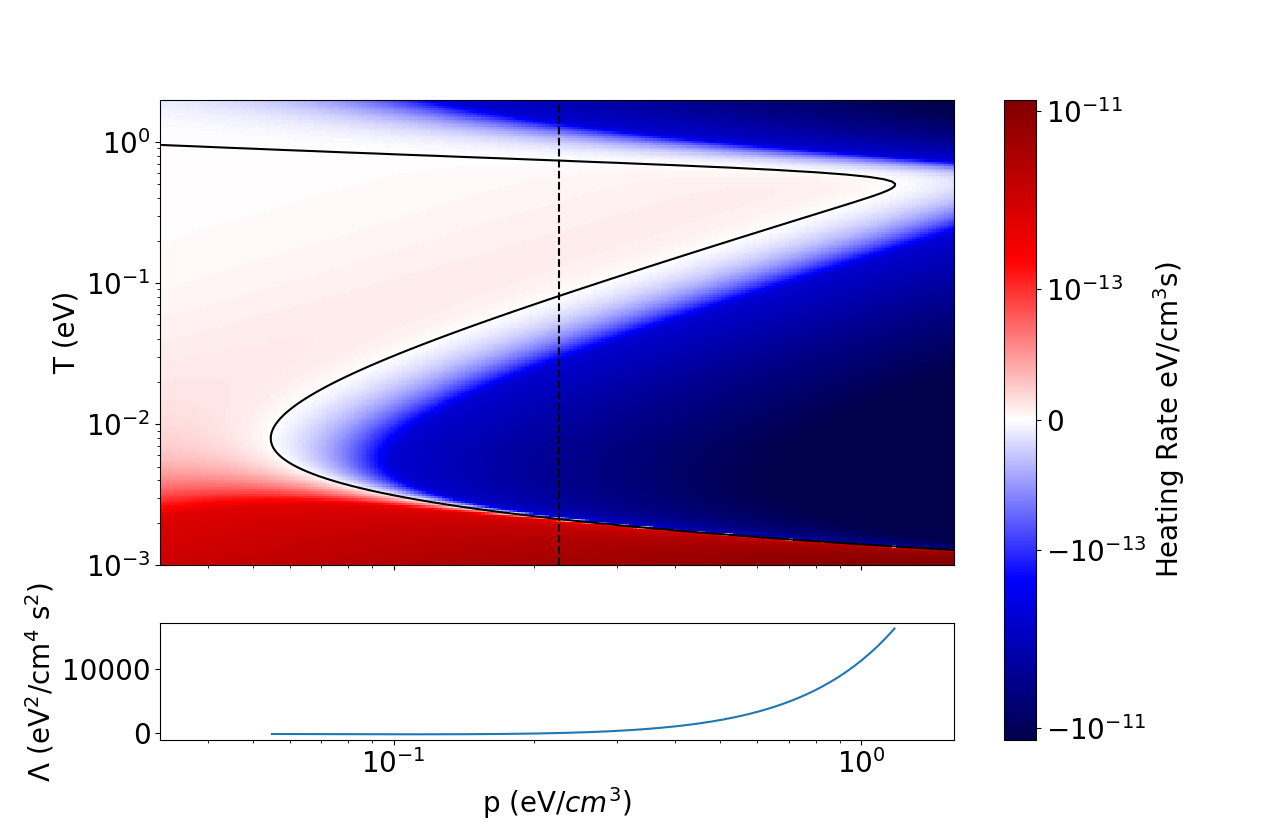}
    \caption{The heating rate, equal to $-\mc L$, given by Eq.~\eqref{eq_mcL_IIK} as a function of pressure and temperature. The bottom panel shows the net cooling function $\kL(p)$.}
\label{fig_cooling_curve}
\end{figure}

As a model for thermal conductivity, we take ${\kappa(p, T) = \kappa_0 T^{1/2}}$ where $T$ is measured in $\mathrm{eV}$. This model is appropriate for a partially ionized plasma, or a neutral gas, in which heat conduction is dominated by neutral particles rather than free electrons. For the equilibrium temperatures visible in Fig.~\ref{fig_cooling_curve}, this is a reasonable model. Following \cite{Parker53}, we adopt for the constant prefactor a value of $\kappa_0 = 1.7\times10^{17} \text{s}^{-1}\text{cm}^{-1}$, corresponding to the thermal conductivity of atomic hydrogen. The characteristic system parameters defined in \S~\ref{sec_scales} take the values given in Table~\ref{table_characteristic_vals}.

\begin{table}[]
    \centering
    \begin{tabular}{c|c}
        $\pcrit$ & $0.23 ~ \mathrm{eV}\mathrm{cm}^{-3}$ \\
        $T_0$ & $0.08 ~ \mathrm{eV}$ \\
        $\mc L_0$ & $5.1\times 10^{-15} ~ \mathrm{eV}~\mathrm{s}^{-1}\mathrm{cm}^{-3}$ \\
        $t_\mathrm{cl}$ & $1.6\times 10^{13} ~ \mathrm{s}$ \\
        $L_0$ & $1\times 10^{15} ~ \mathrm{cm}$
    \end{tabular}
    \caption{Characteristic values for pressure, temperature, cooling function, cooling time, and system size.}
    \label{table_characteristic_vals}
\end{table}

The cooling curve described Eq.~\eqref{eq_mcL_IIK} is shown in Fig.~\ref{fig_cooling_curve} as a function of both pressure and temperature. In the top panel, the blue regions correspond to cooling ($\mc L > 0$) and the red regions correspond to heating. The contour of $\mc L = 0$ is shown in black; the bistable region is the set of pressures that are crossed three times by this curve. The bottom panel shows the net cooling function $\kL(p)$. The critical pressure $\pcrit$, the location where the graph of $\kL(p)$ crosses zero, is shown by the dashed line. For this cooling curve, there is only one critical pressure.  The presence of only one critical pressure guarantees that $\kL_p(\pcrit) > 0$ and so the secondary instability triggered when $\kL_p(\pcrit) < 0$ will not appear.

In its bistable region, this cooling curve satisfies all of the properties assumed by our analysis thus far. Note that $\symbdTdp_{c/h} < 0$ for both stable phases and for all pressures. This property, which is quite general, prevents divergences in Eq.~\eqref{eq_dpdt_firstorder_isolated}.

\subsection{Isolated cloud}

Using the cooling curve defined in Eq.~\eqref{eq_mcL_IIK}, we investigate here the evolution of an isolated cloud and compare to the predictions in \S\ref{sec_evol_isolated_cloud}. At each time step, we evolve the size of every region according to the procedure described in \S\ref{sec_evol_cloudy}.

\begin{figure}
    \centering
    \includegraphics[width=\columnwidth]{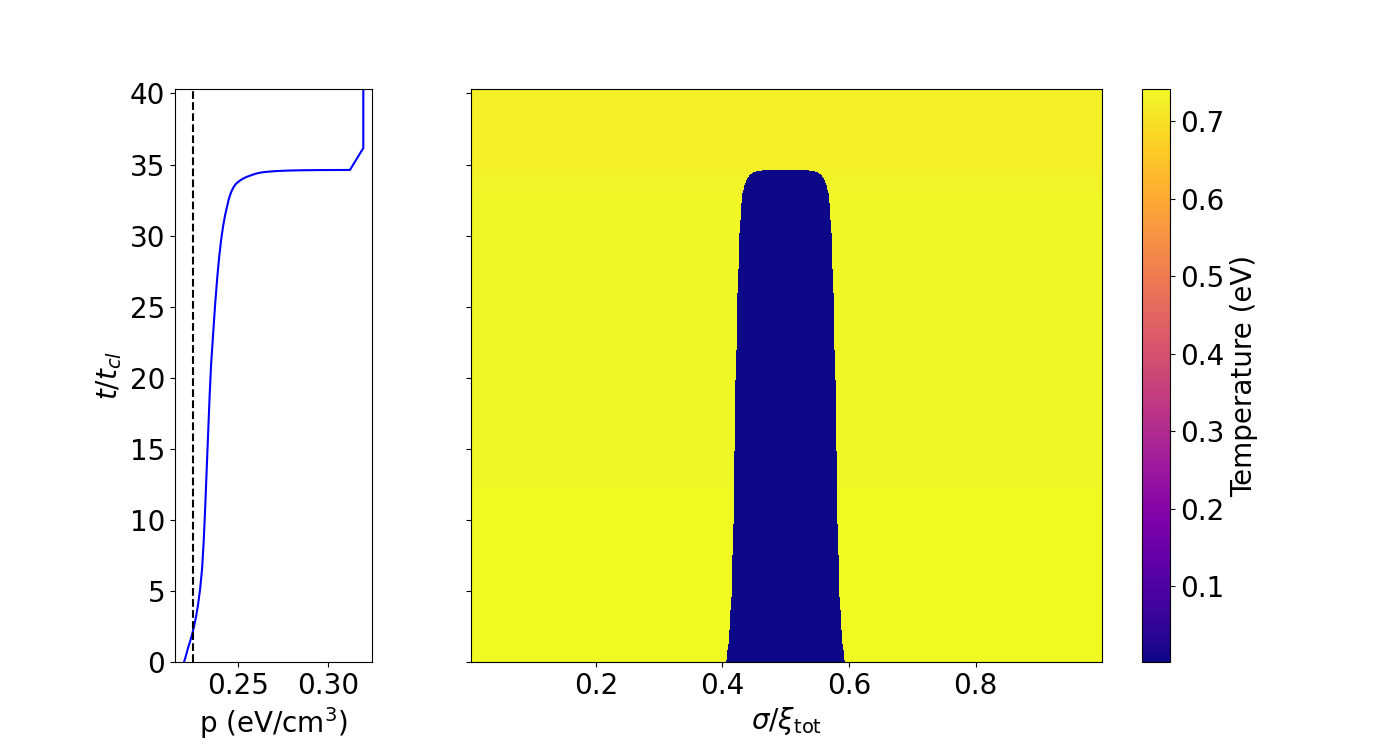}
    \caption{Evolution of a single cold cloud embedded in hot background. In the right panel, each horizontal slice shows the state of the system in Lagrangian coordinates, and the forward time direction is upward.}
\label{fig_evol_cloud}
\end{figure}

For a single cloud of cold material in hot background, under isochoric constraints, Fig.~\ref{fig_evol_cloud} shows a `space-time diagram' of the evolution of the system. In the right panel, each horizontal slice shows the spatial configuration of the system in Lagrangian coordintaes at a fixed time. The vertical axis represents time, with later times placed higher. The left panel shows the pressure as a function of time and the dashed line shows $\pcrit$ for reference. In the case shown, the cloud eventually shrinks as interfront interactions overcome the super-critical pressure. 

\begin{figure}
    \centering
    \includegraphics[width=\columnwidth]{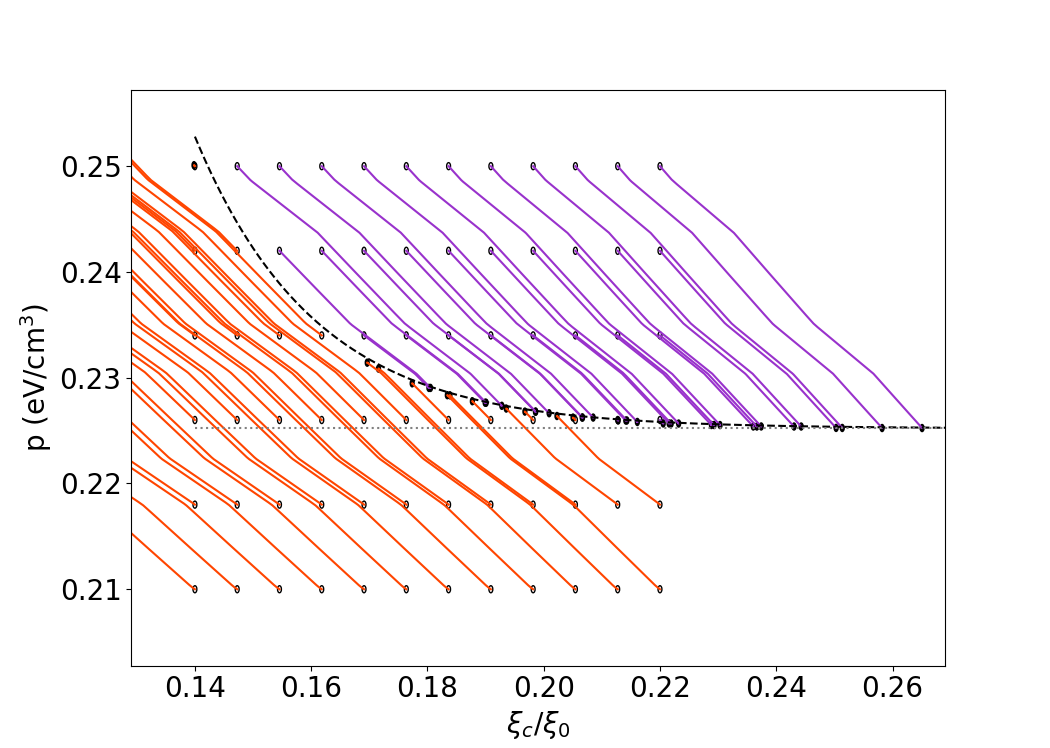}
    \caption{Trajectories of cold clouds. The vertical axis is the system pressure and the horizontal axis is the fraction of total system mass in the cold cloud.}
    \label{fig_trajectories}
\end{figure}

To study the predictions in \S~\ref{sec_evol_isolated_cloud} about equilibrium of isolated clouds, we initialized clouds at a variety of starting conditions and tracked their trajectories over time. The results are shown in Fig.~\ref{fig_trajectories}. The horizontal axis is the fraction of system mass contained in the cold cloud and the vertical axis is the pressure. The hollow circles show starting conditions and the filled black circles show the cloud state at the end of the run. Trajectories for which clouds shrank are shown in red, and those for which clouds grew are shown in purple. The critical pressure is shown by the dotted line and the prediction of Eq.~\eqref{eq_equilib_pprime_Sigmaprime_expanded} is shown by the dashed line.

\subsection{Cloudy  medium}
\label{sec_evol_cloudy}

\begin{figure}
    \centering
    \begin{subfigure}[t]{0.9\columnwidth}
        \includegraphics[width=\columnwidth]{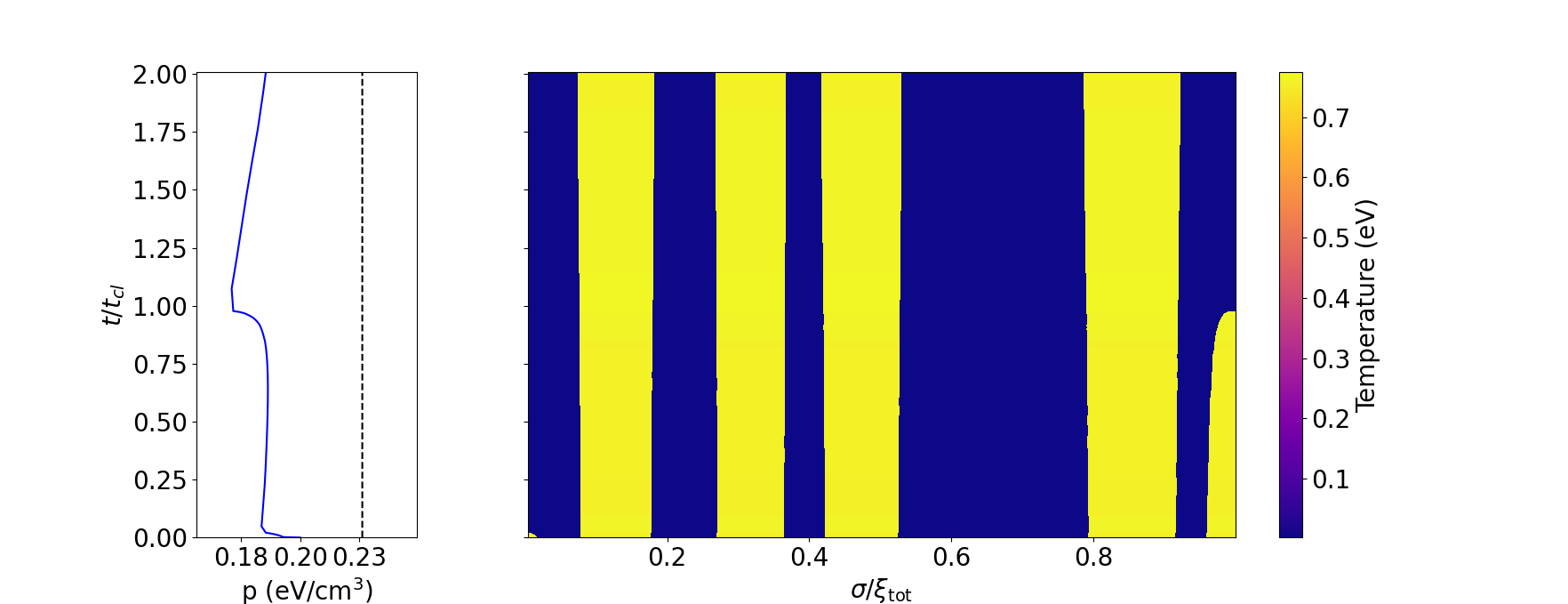}
        \caption{Evolution on short time scales}        
    \end{subfigure}
    \begin{subfigure}[t]{0.9\columnwidth}
        \includegraphics[width=\columnwidth]{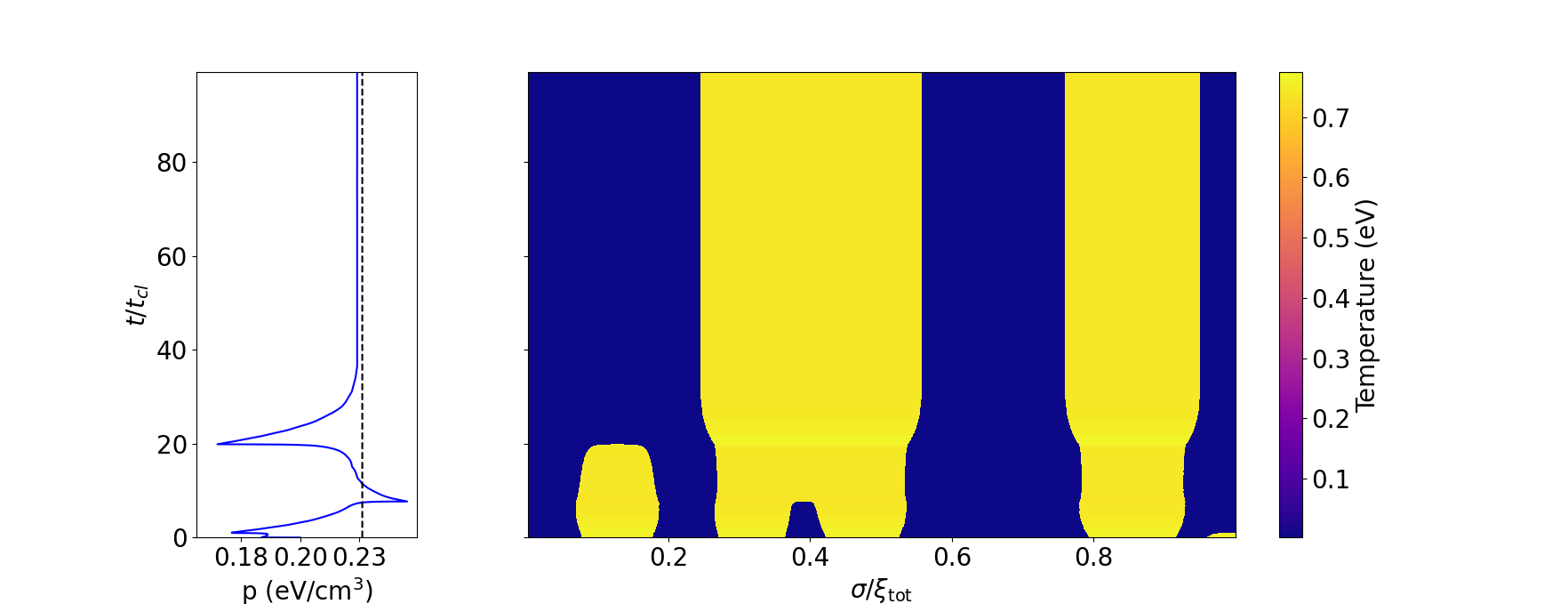}
        \caption{Evolution on intermediate time scales}        
    \end{subfigure}
    \begin{subfigure}[t]{0.9\columnwidth}
        \includegraphics[width=\columnwidth]{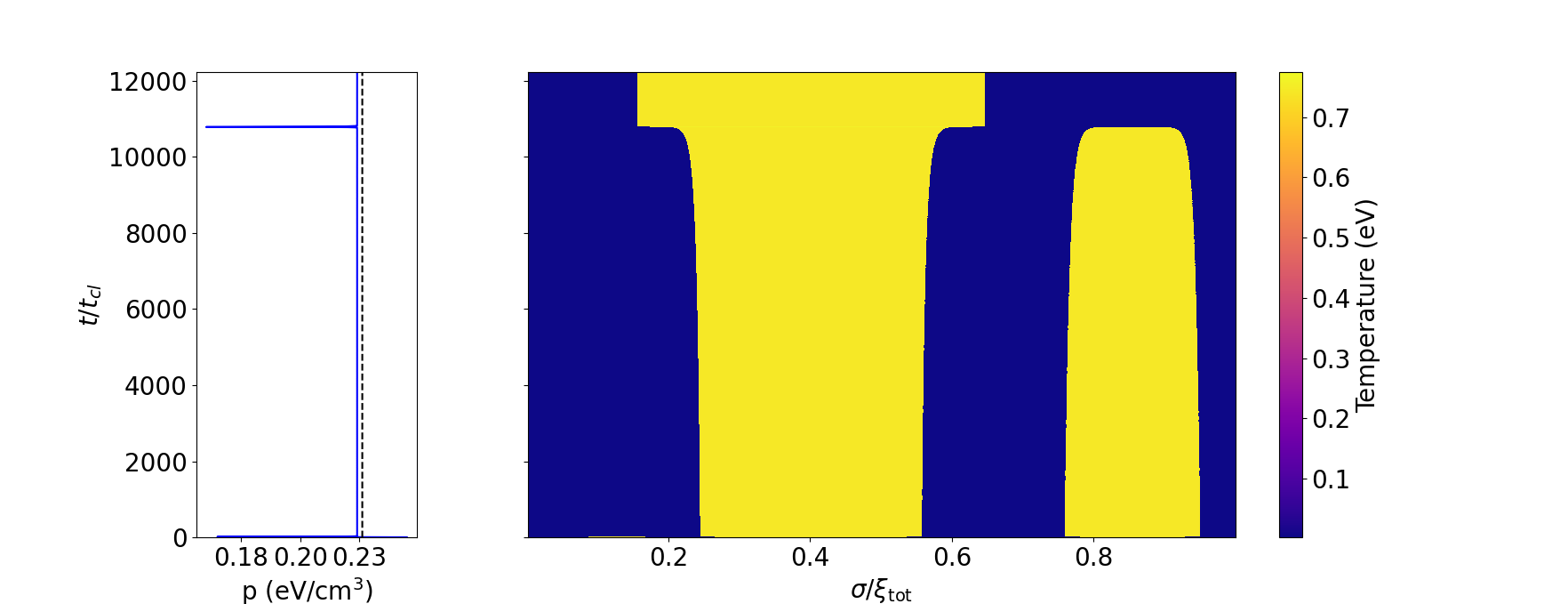}
        \caption{Evolution on long time scales}        
    \end{subfigure}
\caption{Evolution of a random cloudy system shown at multiple zoom levels to highlight various time scales. The vertical axis is time. In the right panel of each subfigure, the horizontal axis represents position in Lagrangian coordinates normalized to the total areal density. In the left panel, the horizontal axis is pressure in units of $\mathrm{eV/cm}^3$.}
\label{fig_evol_medium}
\end{figure}

Finally, we study the evolution of a cloudy medium with complex spatial structure. We initialize a system with twenty regions of alternating phases and random sizes (but fixed total size). The system is then allowed to evolve isochorically by the method described above. The resulting evolution is shown in Fig.~\ref{fig_evol_medium}.

At early times, small clouds rapidly wink out of existence due to interfront interactions (some too quickly to be captured on the plot at all). In the particular case shown, the cold phase initially dominates, with hot regions tending either to merge or to evaporate; of the ten hot regions that were initialized, only two survive beyond about twenty cooling times $t_\mathrm{cl}$. In this process, the pressure rapidly drops. This is consistent with Eq.~\eqref{eq_cloud_lifetime}, which describes the lifetime of an isolated cloud: of ten initial hot regions of random size, the exponential dependence of $t_\mathrm{life}$ on $\Scloud$ means that most hot cloud mass quickly comes to reside in the handful of outliers (clouds that either started large or formed from the merger of two nearby clouds). Note that this picture is approximate as it ignores the effects of other clouds and of non-critical pressure.

When the second-largest hot cloud vanishes, system evolution depends exclusively on the final hot cloud. The evaporation of the second-largest cloud drives the pressure briefly away from critical, but the system rapidly returns to equilibrium by rapid expansion of the remaining cloud. At late times, the system approaches a steady state in which $p$ is close to $\pcrit$. The dynamics discussed in \S\ref{sec_evol_isolated_cloud} and shown in Fig.~\ref{fig_evol_cloud} only apply exactly if the cloud is centered in the region under consideration; if the cloud is closer to one wall, it will tend to migrate toward that wall. However, this process is exponentially slow in the distance to the wall, and is disrupted if the boundary condition that we have assumed at the wall ($q=0$) is violated, i.e. if the wall is not perfectly insulating but instead is allowed to conduct a small amount of heat. For practical purposes, then, even an off-center cloud satisfying Eq.~\eqref{eq_equilib_pprime_Sigmaprime} can be considered to be in steady state.

\section{Discussion}
\label{sec_discussion}

\subsection{The big picture: isobaric and isochoric evolution}
\label{sec_big_picture}

Having described in detail the evolution of a bistable fluid under a variety of conditions, we can now compare the evolution of isobaric and isochoric systems. For both systems, we take as initial conditions a uniform fluid of volume $V$ at pressure $p_0$ and temperature $T_w(p_0)$, the temperature of the unstable (`warm') equilibrium. To this equilibrium, we apply a small, spatially random temperature perturbation. The isobaric system remains at pressure $p_0$ throughout its evolution. The isochoric system remains at volume $V$ and so its pressure varies. For both systems, the pressure remains spatially uniform.

The \textit{isobaric system} evolves as follows:
\begin{enumerate}
    \item On the cooling timescale $t_\mathrm{cl}$, the temperature perturbations grow rapidly.
    \item After a few cooling times, growth saturates at the steady-state temperatures $T_h(p_0)$ and $T_c(p_0)$. The system then consists of a patchwork of hot and cold regions.
    \item System composition evolves via the propagation of conduction fronts, which depends on $\kL(p_0)$ and on the size of each region.
    \item Eventually, the system reaches a homogeneous steady state (all hot or all cold).
\end{enumerate}

The evolution in step 3 is governed by Eq.~\eqref{eq_time_evol_full_sig}. In cases where pressure, rather than interfront interactions, dominates $(\kL/\Q \gg \Gamma)$, cold regions grow if $\kL(p_0) > 0$ (ending in a uniform cold state) whereas hot regions grow if $\kL(p_0) < 0$ (ending in a uniform hot state).

The opposite limit, $\kL/\Q \ll \Gamma$, is most easily analyzed when $p_0 \in \mc P$, in which case $\kL(p_0)$ is identically zero. In this limit, front motion is driven exclusively by interaction with other fronts. This is the case treated by ERS \cite{Elphick_Regev_Spiegel_1991,Elphick_Regev_Shaviv_1992}. As discussed in \S\ref{sec_isobaric_cloudy_medium}, this system evolves toward a uniformly hot or uniformly cold state. Heuristically, the winning phase is determined by the larger of $\hat\Scloud_h$ and $\hat\Scloud_c$ (calculated by summing over all hot and cold regions, respectively). This is only generally true; the nonlinearity of Eq.~\eqref{eq_dSigma_i_dt} renders analysis very difficult and so there are some initial conditions for which the heuristic breaks down. A steady state can be obtained by arranging hot and cold regions periodically, but this state is unstable to most perturbations and can only be reached by meticulously contrived initially conditions; uniform hot and cold states effectively remain, then, the only late-time behaviors worth considering.

When pressure is a dynamical quantity, the system dynamics become considerably richer. Prepared from the same initial conditions, the \textit{isochoric system} evolves as follows:
\begin{enumerate}
    \item On the cooling timescale $t_\mathrm{cl}$, the temperature perturbations grow rapidly. The resultant change in the system's internal energy drives a rapid variation in the pressure.
    \item After a few cooling times, growth saturates at steady-state temperatures $T_h(p)$ and $T_c(p)$, where $p$ is the pressure attained by simultaneous and self-consistent evolution of temperatures and pressure in step 1. The system then consists of a patchwork of hot and cold regions.
    \item System composition evolves via the propagation of conduction fronts, which depends on $\kL(p)$, the size of each region, and in some cases $\partial_t p$.
    \item Concomitantly, pressure evolves as given in Eq.~\eqref{eq_dpdt_chi}. If $p$ is near a critical pressure $\pcrit \in \mc P$ and motion due to interfront interactions is negligible, the following may happen:
    \begin{enumerate}
        \item If $\kL_p(\pcrit) < 0$, then $p$ is rapidly driven away from $\pcrit$. This can be viewed as a secondary instability, the primary being the classical thermal instability. In this case, $p$ will approach a different (stable) critical pressure.
        \item If $\kL_p(\pcrit) > 0$, then $p$ is drawn toward $\pcrit$.
        \item If interfront interactions are not negligible, the situation is more complicated and depends on details of the system configuration. Pressure may approach some $p \neq \pcrit$ at which pressure-driven and interfront contributions to the motion cancel.
    \end{enumerate}
    \item Eventually, the system reaches a steady state, which may be homogeneous or complex in structure, at some pressure $p$. In some cases, $p \notin \mc P$ and instead pressure takes some value determinable only by detailed knowledge of the initial conditions.
\end{enumerate}

The final steady state is determined by the condition that the LHS of Eq.~\eqref{eq_time_evol_full_sig} is zero. Homogeneous configurations are, of course, steady states. Nontrivial steady states are discussed in \S\ref{sec_evol_isolated_cloud} and \S\ref{sec_evol_cloudy}. The simplest such state is a single cloud of either hot or cold phase surrounded by a large region of the other phase. In contrast to the nontrivial steady states in isobaric systems, which require fine tuning and are unstable to perturbations, these isochoric steady states are often stable and accessible through the system's time evolution.

\subsection{Intermediate results}
\label{sec_intermediate_results}

Though the primary result of this work is the description of isochoric evolution outlined above, we highlight here a few of intermediate results obtained in this work that may be of particular interest.

The derivations in \S\ref{sec_isobaric} extend previous work on the motion of conduction fronts (ZP \cite{ZP69}, ERS \cite{Elphick_Regev_Spiegel_1991}, and AMS \cite{Aranson_Meerson_Sasorov_1993}, \textit{inter alia}). The asymptotic approach used here has the advantage of providing a clear dimensionless small parameter to estimate the degree of approximation. To the order of approximation used in this work, the contributions to front motion from different sources add linearly, allowing simultaneous consideration of motion driving by pressure variation and interfront interactions.

Our results, namely Eq.~\eqref{eq_j1_p}, Eq.~\eqref{eq_j1_h}, and the definitions of quantities that appear in these equations, depend only on explicit functionals of the cooling function. These functionals can immediately be applied to any cooling function, provided that a few general properties are satisfied. This has advantages over results given in terms of the conduction front's spatial profile $T(x)$; in general, such a profile must be calculated iteratively (the asymptotic results in this work could, of course, be carried out to higher order through an iterative process, but even in that case, the result would be a general functional applicable to any $\mc L$ with the right properties).



\subsection{Implications for experiments}
\label{sec_experiments}

The results in this work are particularly relevant to systems with fixed volume where the width of a conduction front is not too much less than the system size ($\delta$ is small but not negligibly so). Our results are quantitatively applicable only when the system is effectively one dimensional; in higher dimensions, the curvature of fronts introduces a variety of new effects that complicate the picture \cite{Shaviv_Regev_1994, Meerson_Sasorov_1996}. One important example of such a system is a magnetically confined fusion plasma. In both tokamaks and z-pinches, thermal condensates commonly develop, grow, and propagate. Because the plasma in these systems is magnetized, a one-dimensional treatment becomes a reasonable first approximation. The fact that laboratory fusion plasmas are limited in volume, but typically not constrained by some externally fixed pressure, means that the isochoric case considered in this work is often applicable. Understanding of thermal instabilities, which can provide additional loss channels and degrade confinement, is important for controlled fusion.

Throughout this work, we have assumed that the fluid is everywhere in local thermal equilibrium, but some regimes of interest challenge this assumption. While a major departure from thermal equilibrium would necessitate an entirely different analysis, many relevant systems exist in some kind of quasi-equilibrium and so could be accommodated with minor adaptations.
For instance, laboratory plasmas often exhibit thermal decoupling between ions and electrons and so require treatment with a two-fluid model. If both species are important for heat conduction, the coupled dynamics of the ion and electron fluids can lead to surprising behavior \cite{Jin_Reiman_Fisch_2021}. If the cold phase is dense and cool enough, then non-ideal effects, such as Coulomb coupling in plasma, can become significant. This regime opens new channels for energy exchange between ions and electrons \cite{Fetsch_Foster_Fisch_2023}, which may affect cloud evolution. While accounting for these effects demands a more complicated analysis, thermally bistable plasma could prove to be an advantageous platform for studying such fundamental effects.

The secondary instability found in \S~\ref{sec_isochoric_isolated_front} can drive a rapid change in pressure and a rearrangement of hot and cold regions. This has the potential to yield a detectable signal in an experiment probing thermally bistable fluid.

Some astrophysical and space plasmas, such as solar prominences, are relevant to the cases considered in this work for reasons similar to those given above. A topic of intense interest in astrophysics is the study of matter at extreme pressures and densities \cite{Bailey_Nagayama_Loisel_Rochau_Blancard_Colgan_Cosse_Faussurier_Fontes_Gilleron_et_2015, García_Fabian_Kallman_Dauser_Parker_McClintock_Steiner_Wilms_2016, García_Kallman_Bautista_Mendoza_Deprince_Palmeri_Quinet_2018}. Though such materials may or may not be photoionized in nature, a leading technique for probing opacity in these regimes is the photoionization of a sample using an intense burst of radiation, often generated by a z-pinch or laser-driven implosion \cite{Rochau_Bailey_Falcon_Loisel_Nagayama_Mancini_Hall_Winget_Montgomery_Liedahl_2014, Mancini_Lockard_Mayes_Hall_Loisel_Bailey_Rochau_Abdallah_Golovkin_Liedahl_2020, Heeter_Perry_Johns_Opachich_Ahmed_Emig_Holder_Iglesias_Liedahl_London_et_2018, Opachich_Heeter_Johns_Dodd_Kline_Krasheninnikova_Mayes_Montgomery_Winget_Urbatsch_et_2022}. If spatial inhomogeneities develop in the photoionized plasma and have lifetimes comparable to the timescale of the experiment, opacity measurements can be drastically affected. In this work, we have shown that spatial inhomogeneities survive in thermally bistable plasma even at late times. For photoionized plasma experiments in a bistable regime, the persistence of these spatial structures may be an important consideration.

\section*{Acknowledgments}

This work was supported by the Center for Magnetic Acceleration, Compression, and Heating (MACH), part of the U.S. DOE-NNSA Stewardship Science Academic Alliances Program under Cooperative Agreement DE-NA0004148 and by the National Science Foundation under Grant No. PHY-2308829.

\appendix 

\section{Supplemental Notation}
\label{sec_appendix_definitions}

In \S\ref{sec_isobaric_interacting_fronts} we considered a hot cloud of finite size, and in the course of the derivation we defined $\lambda_h, k_h,$ and $\Theta_h$. Their counterparts for cold regions are straightforward to derive by the same process, but care needs to be taken with signs and so we omitted the definitions from the main text. The definitions are as follows:
\begin{equation}
\label{eq_lambda_c_k_c_Theta_c_def}
\begin{split}
    \lambda_c &\doteq \left(\frac{\partial}{\partial T}\kappa \mc L\right)\Big|_{T_c} ,
    \\
    k_c &\doteq \frac{\sqrt{\lambda_c}}{n_c \kappa(T_c)} ,
    \\
    \Theta_c &\doteq \int_0^{T_w-T_c} d\tau' \left[\frac{k_c n(T_c+\tau')\kappa(T_c+\tau')}{\sqrt{2\int_0^{\tau'} d\theta \kappa(T_c + \theta)\mc L(T_c+\theta)}} - \frac{1}{\tau'}\right] .
\end{split}
\end{equation}

\bibliography{photoion}

\end{document}